\documentclass[11pt]{article}

\usepackage{braket}
\usepackage{amsmath}
\usepackage{amsfonts}
\usepackage{geometry}
\usepackage{mathtools}
\usepackage{amssymb}
\usepackage{esint}
\usepackage{cancel}

\usepackage{subcaption}
\DeclareCaptionFormat{custom}
{\textbf{#1#2}\textit{\small #3}}
\def\hh{\hspace{0.5mm}}

\def\dd{\mbox{d}}

\title{Renormalisation for Reaction-Diffusion Systems with Non-Local Interactions}
\author{Chris D Greenman}

\begin{document}
\maketitle


\begin{abstract}
Models of reaction diffusion processes usually employ discrete lattice models with particles interacting at the same site, resulting in localized reactions in the continuum limit. Here, various non-local interactions are considered, and two features reported. Firstly, it is shown that sufficiently non-local interactions will regulate ultra-violet divergences that perturbative methods with local interactions produce. However, in asymptotic regimes, infra-red divergences persist and ultra-violet divergences can reappear. Renormalisation methods are shown to report the same universal behaviour as local interactions at critical points. Secondly, the renormalisation group can be interpreted as a space-time-field rescaling that preserves action structure. This can be used to extract solutions to Callan-Symanzik equations directly without having to solve (or construct) the equation. These observations are exemplified for two paradigm models; annihilation $A_p+A_q\rightarrow \phi$, and this process paired with branching, birth and death.
\end{abstract}


\section{Introduction}

Reaction-diffusion processes are an enveloping term describing systems with discrete, mobile interacting entities \cite{Liggett1985}. These can be used to describe systems on many scales, ranging from molecular reactions \cite{Doi1976, Doi1976b} to spatial predator-prey systems \cite{Dobramysl2018}, for example. Although the objects in such populations differ substantially in nature, they shall be referred to as \emph{particles} from this point forward. Techniques to investigate such models can be quite varied, including semigroup approaches \cite{Banasiak2006, Goldstein2017}, stochastic approaches \cite{Gardiner1985, Liggett1985}, sample size methods \cite{Van1992}, and field theory techniques \cite{Tauber2014} to name a few. 

It proves useful to describe the interactions algebraically. A reaction written as $A + A \rightarrow \phi$, for example, would indicate either pairs of colliding particles that annihilate, or pairs that combine into a different class, say $B$, which are not of interest and subsequently ignored. This is a paradigm model for such systems \cite{Doi1976, Doi1976b, Lee1994}. This also has a natural extension where branching is also included $A \rightarrow A + A$ \cite{Cardy1998}. Spatial effects with some form of diffusion is usually implicit with these models. Such processes can contain critical points of phase transition. In particular, for pure annihilation, the asymptotic large time behaviour exhibits a qualitative change at critical spatial dimension $d_c = 2$. If branching is added to the process, a different critical behaviour emerges; the point at which branching and annihilation rates conspire to produce a non-zero steady state has critical dimension $d_c=4$. Renormalisation methods of field theory have proven useful for examining criticality of such systems \cite{Lee1994, Peliti1986, Tauber2014} and is the subject of this study. 

Usually, these methods are developed with the aid of path integral techniques. These often start with an underlying lattice model \cite{Lee1994, Peliti1985} such that pairs of particles can only annihilate or spawn on single lattice sites. Such systems can then limit to a continuum process, with a `lattice constant' $a$ (indicating separation between proximal sites) being reduced to zero. However, for many systems of interest, either because of particle size, or interactions acting over a distance, such local models are insufficient to capture this behaviour. For example, the original work of Doi \cite{Doi1976, Doi1976b} considered non-local interactions (albeit without path integrals) for the annihilation model $A_p + A_q \rightarrow \phi$, where particles at positions $p$ and $q$ annihilate when sufficiently close to each other. If $p$ instead represents the age of a particle (rather than position), a birth process can be written as $A_p \rightarrow A_p+A_0$, where a new offspring with age zero appears. This is inherently non-local \cite{Greenman2017}, and is also an example of stochastic resetting which has recently generated significant interest \cite{Evans2020}. Dealing with non-local interactions has also been considered in the guise of polymer reactions \cite{DeGennes1982, Kleinert2006, Kleinert2006b}. In some cases the polymer is a string and interactions occur during contact (so in some sense is still local), but in some cases, the polymers have width and the interactions are non-local. Adapting renormalisation methods to deal with such non-local behaviour is thus of interest and the primary aim of this work.

Utilising renormalisation techniques to capture the asymptotic behaviour of the systems generally encompasses two features \cite{Lee1994, Tauber2014}. Firstly, perturbation expansions in field theory often involve `ultra violet' (UV) divergent terms, which renormalisation methods can systematically remove (albeit just for renormalisable theories). Secondly, the asymptotic behaviour of interest often involve `infra red' (IR) divergences in critical regions of interest, and renormalisation group techniques are used to connect features in these divergent domains near critical points to perturbative terms arising from convergent regions. 

In this work, path integral renormalisation methods for non-local interactions are considered in detail. Specifically, we show that non-local interactions have the natural effect of regulating UV divergences, which either have reduced divergence, or no longer diverge below the critical dimension of interest. However, the IR divergences still exist and need to be treated to examine asymptotics. The non-local nature of interactions can make the perturbative re-summations needed for renormalisation somewhat intractable. However, the rescaling that underlies the renormalisation group will be seen to have the convenient effect of converting non-local into local interactions. Subsequently, the critical dimensions and associated universal properties observed for locally interacting systems, that demarcate a dimensional transition in asymptotic decay exponents, are preserved. We show this for both the annihilation process \cite{Lee1994} and when branching, spontaneous birth and death is included \cite{Cardy1998}. 

Intrinsic to these methods are the Callan-Symanzik equations used to connect features at different scales. Or, more specifically, the solutions to these equations run along partial differential equation characteristics, which connect features in regions that are perturbatively divergent (typically as one approaches a critical point) to the same feature in a convergent region. In this work, we point out that a space-time-field rescaling can be implemented, conditional on a preservation of action structure, that extracts the same solutions directly, without having to construct or solve these equations.

In the next section we introduce field theoretic machinery used to formalise non-locally interacting models. This includes a brief introduction to Doi-Peliti methods, a Hubbard-Stratonovich transform that can conveniently disentangle the non-local interactions at the expense of additional fields. The third section introduces renormalisation methods for non-local annihilation processes, where it is shown that Callan-Symanzyk solutions are extractable via a space-time-field rescaling without need of the equations. After establishing both UV and IR divergences emerge asymptotically (for non-local processes), it is shown how rescaling reduces non-local interactions to local behaviour asymptotically, and the same  critical universality results. The fourth section adapts these methods to a more general process involving annihilation, branching, birth and death. Conclusions complete the work. 


\section{Field Theory}

This section contains the following. Firstly, some conventions are specified. Then, to fully define the models of interest, the non-local nature of some interactions are specified. Doi-Peliti methods for constructing path integrals are then introduced. This is done without recourse to lattice methods and is more akin to Doi's original continuous approach \cite{Doi1976, Doi1976b}. The non-local nature of interaction leads to path integral actions that have double integrals over space, which are awkward to deal with. A Hubbard-Stratonovich transform is then shown to disentangle this, at the price of two extra fields. Investigation of mean field behaviour completes this section, providing the necessary background for subsequent renormalisation methods. 


\subsection{Conventions}

For the sake of clarity, we briefly note the following. Firstly, the following shorthand is adopted, where $d$ represents the spatial dimension:
\begin{equation}
\int_t \equiv \int_{\mathbb{R}} \dd t,
\hspace{0.5cm}
\int_p \equiv \int_{\mathbb{R}^d} \dd p,
\hspace{0.5cm}
\fint_\omega \equiv \frac{1}{2\pi}\int_{\mathbb{R}} \dd \omega,
\hspace{0.5cm}
\fint_k \equiv \frac{1}{(2\pi)^d}\int_{\mathbb{R}^d} \dd k,
\hspace{0.5cm}
\textrm{and}
\hspace{0.5cm}
\cancel{\delta}(k) \equiv (2\pi)^d\delta(k).
\nonumber
\end{equation}
Secondly, Fourier transforms are defined as $f(p,t)=\fint_{\omega}\fint_k e^{i(pk-\omega t)}f(k,\omega)$. Thirdly, propagators in momentum space are understood in the sense that, for example, $G_{\bar{\psi}\psi}(k,l)=\delta(k+l)G_{\bar{\psi}\psi}(l)$ (see Fig. \ref{FeynmanPic} or Eq. (\ref{AnnProp}) for details). 


\subsection{Interaction Profiles}

\begin{figure}[t!]
\setlength{\unitlength}{0.1\textwidth}
\begin{picture}(8,1.8)
\put(0,0){\includegraphics[width=14cm]{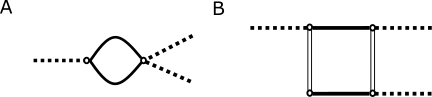}}
\put(1.85,1.20){$G(-k)$}
\put(1.85,0.00){$G(k)$}
\put(1.50,0.40){$R$}
\put(2.55,0.40){$R$}
\put(5.90,1.00){$G(-k)$}
\put(5.95,0.20){$G(k)$}
\put(5.10,0.60){$R(k)$}
\put(6.95,0.60){$R(-k)$}
\end{picture}
\centering
\caption{A) Single loop Feynman diagram for the annihilation model in \cite{Lee1994} with local interaction rate $R$. B) Corresponding diagram for non-local interaction function $R(k)$ (in momentum representation). Dotted lines indicate truncated propagators. In both cases external momenta are zero and the loop momentum is $k$.}
\label{FeynComp}
\end{figure}
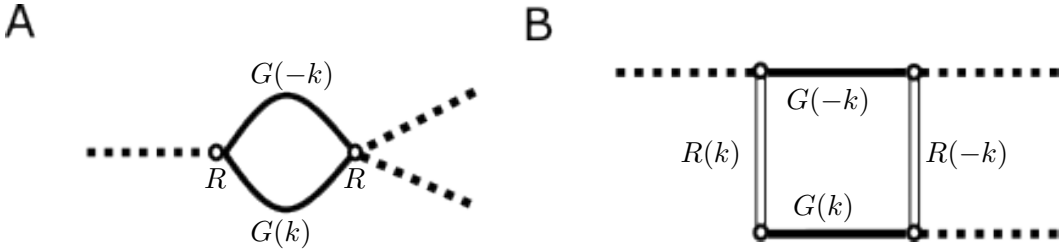

Before describing some possible interaction profiles, we briefly summarise the different effects that local and non-local models of interaction have on Feynman diagrams. In Fig. \ref{FeynComp}A we see a divergent diagram arising in the pure annihilation model of \cite{Lee1994} with local interaction. This diagram has two constant interaction terms $R$, and two propagators, each with a quadratic denominator in momentum space (see \cite{Lee1994} or Eq. (\ref{AnnProp}) for details). Integration over momentum and frequency (dimension $d+2$) thus gives a superficial degree of divergence of $d-2$. Once non-local effects are enabled, the interaction constants $R$ becomes functions $R(k)$ in momentum space (Fig. \ref{FeynComp}B). Depending on the form of interaction (see below), these conspire to further reduce the degree of divergence and diagrams are less divergent for non-local interactions. For example, if $R(k)$ has a quadratic denominator, the degree of divergence becomes $d-6$, whereas if $R(k)$ decays exponentially, the diagram is convergent in any dimension. Both cases arise for specific examples of interest described below.

Next, then, a range of possible interactions are considered. In general, we have an interaction rate $R_{pq}\equiv R(|p-q|)$ that gives the instantaneous rate of reaction between two separate particles positioned at loci $p$ and $q$. It is assumed that this interaction is isotropic and homogeneous, so in all cases it is just a function of separation distance $|p-q|$. The different forms considered are highlighted in Table \ref{Interactions}. 

\begin{table}[t!]
\renewcommand{\arraystretch}{2}
\centering
\begin{tabular}{ |l | c | c | c |}
\hline 
Interaction & $R(r)$ & $R(k)$ & $\textrm{Var}\left(\frac{R(r)}{R}\right)$ \\ \hline \hline
Local & $R\delta(r)$ & $R$ & 0 \\ \hline
Normal & $R\left(\frac{\lambda^2}{\pi}\right)^{\frac{d}{2}}e^{-\lambda^2 r^2}$ & $Re^{-\frac{k^2}{4\lambda^2}}$ & $\frac{1}{2\lambda^2}$ \\ \hline 
Screened Poisson & $R(2\pi)^{-\frac{d}{2}}\lambda^2\left(\frac{\lambda}{r}\right)^{\frac{d}{2}-1}K_{\frac{d}{2}-1}\left(\lambda r\right)$ & $R\frac{\lambda^2}{\lambda^2+k^2}$ & $\frac{2}{\lambda^2}$ \\ \hline   
Spherical & $\frac{R\lambda^d}{V_d}\theta(|r|<\frac{1}{\lambda})$ & $R(\frac{2\lambda}{k})^{\frac{d}{2}}\Gamma(\frac{d}{2}+1)J_{\frac{d}{2}}(\frac{k}{\lambda})$ & $\frac{1}{\lambda^2(d+2)}$ \\ \hline
Riesz Potential & $\frac{R}{N_{f,\nu}c_\nu}\int_y\frac{f(y)}{(r-y)^\nu},(0<\Re(\nu)<d)$ & $\frac{R\hat{f}(k)}{N_{f,\nu}k^{\nu}}$ & -
\\ \hline
\end{tabular}
\caption{A range of interaction profiles $R(r)=R\hat{R}(r)$ that are a function of separation $r$. The parameter $\lambda$ in all cases represents a form of precision. The function $J_{\frac{d}{2}}$ is a Bessel function of the first kind. The function $K_{\frac{d}{2}-1}$ is a modified Bessel function of the second kind. The term $V_d = \frac{\pi^{\frac{d}{2}}} {\Gamma(\frac{d}{2}+1)}$ is the volume of a $d$ dimensional unit sphere. The Reisz potential is defined provided the test function $f$ decays sufficiently quickly. The value $c_\nu = \pi^{\frac{d}{2}}2^\nu\frac{\Gamma(\frac{\nu}{2})}{\Gamma(\frac{d-\nu}{2})}$ is a Fourier transform normalisation term, and $N_{f,\nu}$ is a normalisation constant to ensure a total integral of $R$.}
\label{Interactions}
\end{table}

The \emph{Local} form $R(r) = R\delta(r)$ is the profile used for most lattice models \cite{Lee1994, Peliti1985}, with interactions occurring on individual lattice sites. Note that this takes the constant form $R(k)=R$ in momentum space, contributing no terms of consequence when ultra-violet divergence is considered. All the remaining cases considered are non-local, with a form $R(r)=R\hat{R}(r)$ involving an overall rate $R$ multiplied by a normalised probability density $\hat{R}(r)$ that is a function of separation, $r$ (apart from the Reisz potential described below). The \emph{Normal} case gives a natural form of interaction, with a strongly decaying function in momentum space, which will be seen to provide a strong form of regulation for UV divergence. The \emph{Screened Poisson} distribution generalizes the Coulomb potential into a normalizable function, satisfying an equation of the form $(\nabla^2-\lambda^2)\hat{R} = -\delta(r)$. This structure also has a convenient Fourier transform $\frac{\lambda^2}{\lambda^2 + k^2}$ which will prove pliable when Feynman integrals in momentum space are considered. Note the quadratic nature of the denominator, which will be seen to provide some regulation when ultra-violet divergences are considered. The \emph{Spherical} profile is perhaps the most natural. This can be interpreted as giving particles a finite, specific width, with interactions occurring at fixed rates when particles touch. It can also be interpreted as interactions between point particles that require a proximity threshold. This is the form considered with field theory by Doi \cite{Doi1976, Doi1976b}, originally considered as interaction on contact by Smoluchowski \cite{Von1917} and later as an interaction rate by Teramoto and Shigesada \cite{Teramoto1967}. Although the spherical model is a natural choice, the hard edges of the interaction profile do not produce a very forgiving function in momentum space. However, asymptotically, this takes the form $\mathcal{O}\left({k^{-\frac{d+1}{2}}}\right)$ and so provides weaker regulation than the screened Poisson for $d\simeq 2$. Finally, the \emph{Reisz potential} can provide varied regulation, in particular, it can provide regulation between the delta function (none) and the screened Poisson (quadratic), although analytic results will only be possible for certain choices of test function $f$. 

All the non-local cases except Reisz have some form of precision parameter $\lambda$ such that as $\lambda \rightarrow \infty$ we recover the local interaction. 


\subsection{Doi-Peliti Methods}

Here the requisite Doi-Peliti field theory is developed. This material can be found elsewhere and is just an overview. Note that the original development by Doi \cite{Doi1976, Doi1976b} does not use a discrete lattice, but is entirely continuous in nature. It does not treat path integrals, however, which are needed. Many of the path integral formulations seen today \cite{Peliti1985, Tauber2014} utilise lattice formulations. The description below does not, and directly implements a continuous path integral framework \cite{Greenman2017}. Although this offers an alternative derivation and a point of interest, being more closely aligned with Doi's original approach, the end results are the same as those of lattic methods.

Firstly, then, the creation and annihilation operators $a_q^\dag$ and $a_p$ are introduced via the commutation relations,
\begin{equation}
[a_p,a_q^\dag] = \delta(p-q),
\hspace{0.5cm}
[a_p,a_q] = 0,
\hspace{0.5cm}
[a_p^\dag,a_q^\dag]=0,
\nonumber
\end{equation}
where $p,q$ represent positions in space.

Next we introduce fundamental $n$-particle states via $\ket{\mathbf{p}_n} = a_{p1}^\dag\dots a_{p_n}^\dag\ket{\phi}$, where $\ket{\phi}$ is the vacuum (zero population) state. Using the commutation relations yields the normalisation,
\begin{equation}
\braket{\mathbf{p}_n|\mathbf{q}_m} = \delta_{mn} \sum_{\pi \in \mathcal{S}_m} \prod_{i=1}^m \delta(p_i-q_{\pi(i)}),
\nonumber
\end{equation}
where $\mathcal{S}_m$ denotes the $m^\mathrm{th}$ symmetry group. 

Now, the state of the system $\ket{\chi_t}$ is introduced as a probabilistic smearing across fundamental states, where,
\begin{equation}
\ket{\chi_t} = \sum_n\int \dd \mathbf{p}_n \hh \frac{f_n(\mathbf{p}_n,t)}{n!}\ket{\mathbf{p}_n}.
\nonumber
\end{equation}
Here $f_n(\mathbf{p}_n,t)\dd \mathbf{p}_n$ represents the probability that the population size is $n$ and there exists a labelling such that the $i^\mathrm{th}$ particle can be found inside the hypercube from $[p_i,p_i+\dd p_i]$. 

To proceed further, we require coherent states. For a complex function $\zeta(p)$ of space, the coherent state $\ket{\zeta}$ is defined as,
\begin{equation}
\ket{\zeta} = e^{\int_p \zeta(p) a_p^\dag}\ket{\phi}.
\nonumber
\end{equation}
These states normalise via $\braket{\zeta|\zeta'} = \exp\int_p\bar{\zeta}\zeta'$ and are utilised in three fundamental ways. 

Firstly, the initial state of the system is assumed to be composed of a Poisson number of particles with mean $m$ such that each particle occupies a position $p$  with probability density $\omega(p)$. In this situation, the initial state is given by,
\begin{equation}
\ket{\chi_0} = e^{-m}\ket{m\omega}. 
\nonumber
\end{equation}

Secondly, the unit coherent state $\bra{1}$ is required to calculate moments, where $1$ denotes the constant function. In particular, for the zeroth moment, $\braket{1|\chi_t}=1$ represents total probability conservation. 

Thirdly a resolution of identity is required to construct path integrals, where we find that,
\begin{equation}
I = \iint \frac{\mathcal{D} \psi \hh \mathcal{D} \bar{\psi}}{2\pi}e^{-|\psi|^2}\ket{\psi}\bra{\bar{\psi}}
=\iint \frac{\mathcal{D} [\operatorname{Re}\psi] \hh \mathcal{D} [\operatorname{Im}\psi]}{\pi}e^{-|\psi|^2}\ket{\psi}\bra{\bar{\psi}}.
\nonumber
\end{equation}
A slightly different formulation is used in \cite{Greenman2017}, which results in path integrals over spaces of real functions, rather than two conjugate complex fields.

To build a stochastic model, we start with a suitable Liouvillian operator to capture the dynamics of interest. To do this the associated models need to be specified. The first, Model I, is a pure annihilation model, and is a non-local version of the model considered by Lee \cite{Lee1994}. The second model generalizes this, and includes processes of branching, spontaneous birth and death, giving a non-local version of the model considered by Janssen \cite{Janssen1999}. Both models assume an underlying diffusion at rate $D$, and are summarized below:
\begin{align}
\text{Model I:} & \text { Annihiliation} &
\text{Model II:} & \text { Full} \nonumber \\
A_p + A_q & \xrightarrow[]{R_{p-q}} \phi & A_p + A_q & \xrightarrow[]{R_{p-q}} \phi \nonumber\\
&& A_p & \xrightarrow[]{Q_{p-q}} A_p + A_q 
\label{ModDes}\\
&& A_p & \xrightarrow[]{M_p} \phi \nonumber\\
&& \phi &\xrightarrow[]{B_p} A_p \nonumber
\end{align}
For the full Model II, then, the Liouvillian would take the form,
\begin{align}
\mathcal{L} & = \int_p Da_p^\dag \nabla^2 a_p 
+\int_p\int_q R_{pq}\left(a_pa_q-a_p^\dag a_q^\dag  a_pa_q \right)
+\int_p\int_q Q_{pq}\left(a_p^\dag a_q^\dag a_p-a_p^\dag a_p \right) \nonumber \\
& + \int_p M_p\left(a_p-a_p^\dag a_p \right)
+ \int_p B_p\left(a_p^\dag -1 \right).
\label{LiouvFull}
\end{align}  
The terms, respectively, represent diffusion, annihilation, branching, death and spontaneous birth. The Liouvillian for pure annihilation (Model I) would thus just retain the first two integrals.

The dynamic equation (for any Liouvillian) is given by $\ket{\psi_t} = e^{\mathcal{L}t}\ket{\psi_0}$. We can then examine features of interest. For example, the particle density is found from,
\begin{equation}
X(p,t) = \braket{1|a_p|\chi_t} = \braket{1|a_pe^{\mathcal{L}t}|\chi_0}
= \mathcal{N}^{-1}\int \mathcal{D}[\psi_{p,t}]\mathcal{D}[\bar{\psi}_{p,t}]\psi_{p,t}e^{S},
\label{MeanDen}
\end{equation}
where the last term utilises the resolution of identity to form a path integral via the usual methods of time slicing \cite{Greenman2017, Peskin1995, Sredniki2007, Tauber2014}, which converts the Liouvillian operator into an action $S$. The normalisation constant $\mathcal{N} = \int\mathcal{D}[\psi_{p,t}]\mathcal{D}[\bar{\psi}_{p,t}]e^{S}=1$ is just the total probability density and could be ignored at this point. However, Jacobians in corresponding path integral measures will appear during scaling transformations, which will subsequently cancel if the term $\mathcal{N}$ is included. Implementing a standard Doi field shift $\bar{\psi}_p \rightarrow \bar{\psi}_p+1$ finally results in the following action (for the full Model II),
\begin{eqnarray}
S & = & -\int_t\int_p \bar{\psi}_p(\partial_t-D\nabla^2)\psi_p
-\int_t\int_p\int_q R_{pq}\left( \bar{\psi}_p\psi_p\left[\psi_q+\frac{1}{2}\bar{\psi}_q\psi_q\right] \right)
+n_0\int_p \bar{\psi}_{p}(t=0)
\nonumber \\
&& +\int_t\int_p\int_q Q_{pq}\left( \bar{\psi}_p\left[\bar{\psi}_q+1\right]\psi_q \right)
-\int_t \int_p M_p \bar{\psi}_p \psi_p
+\int_t \int_p B_p \bar{\psi}_p.
\label{ActionFull}
\end{eqnarray}
The pure annihilation Model I just retains the first line in this equation. Note that the time dependences of the fields are now suppressed for simplicity of notation (the parameters are taken to be time independent). There is also an important simplification here. Specifically, we assume that the initial spread of particles is uniform. That is, we take initial density $\omega(p) = \frac{1}{V}$ per particle over some large volume $V$. As the mean initial number of particles is Poisson with mean value $m$, the corresponding term in the action then becomes $m\int_p \omega_p\bar{\psi}_{p}(t=0) = \frac{m}{V}\int_p\bar{\psi}_{p}(t=0)$. Defining the value $n_0 = \frac{m}{V}$ as the particle density then reproduces the corresponding term in the action above. Note that if we implement the local form of interaction $R_{pq}= R\delta(p-q)$, remove branching ($Q_{pq} = 0$), spontaneous birth ($B_p=0$) and death ($M_p=0$),  then the site specific local annihilation action seen in \cite{Lee1994} is recovered.

Now, for non-local interactions, perturbative expansions using Eq. (\ref{ActionFull}) will involve integrals over two space variables. This will implicate two nodes in corresponding Feynman graphs. Although this in inconvenient, it can be formally addressed via a Hubbard-Stratonovich transform. 


\subsection{Hubbard-Stratonovich Transform}

\begin{figure}[t!]
\setlength{\unitlength}{0.1\textwidth}
\begin{picture}(9.5,4)
\put(0,0){\includegraphics[width=16cm]{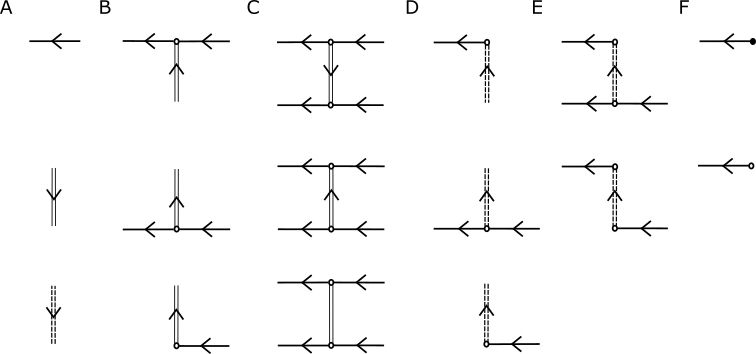}}
\put(0.55,3.45){$G_{\bar{\psi}\psi}$}
\put(0.35,3.85){{\tiny $\psi$}}
\put(0.85,3.85){{\tiny $\bar{\psi}$}}
\put(0.75,1.85){$R$}
\put(0.60,1.40){{\tiny $v$}}
\put(0.60,2.35){{\tiny $u$}}
\put(0.80,0.42){$Q$}
\put(0.60,-0.02){{\tiny $\hat{v}$}}
\put(0.60,0.90){{\tiny $\hat{u}$}}
\put(2.02,3.88){{\tiny $-1$}}
\put(2.08,1.30){{\tiny $\frac{1}{2}$}}
\put(2.08,-0.05){{\tiny $1$}}
\put(4.08,3.35){$-\frac{1}{2}R(k)$}
\put(4.08,1.85){$-\frac{1}{2}R(k)$}
\put(4.08,0.45){$-\frac{R(k)+R(-k)}{2}$}
\put(5.82,3.88){{\tiny $1$}}
\put(5.82,1.30){{\tiny $1$}}
\put(5.82,-0.05){{\tiny $1$}}
\put(7.58,3.35){$Q(k)$}
\put(7.58,1.85){$Q(k)$}
\put(9.15,3.70){$n_0$}
\put(9.15,2.20){$B(k)$}
\end{picture}
\centering
\caption{Feynman perturbative diagrams for action post Hubbard-Stratonovich transforms. A) Three classes of propagator. B) Three vertices in perturbative expansion for annihilation terms, with coefficients. C) Annihilation node pairs connected by an annihilation propagator $R$ with momentum $k$ in each direction (each with coefficient $-\frac{1}{2}$) correspond to a single, directionless propagator (with coefficient $-\frac{R(k)+R(-k)}{2}$). D) Three vertices in perturbative expansion for branching terms (all have coefficient $1$). E) Possible node pairs for branching propagator $Q$. F) Terminal nodes. The $n_0$ (filled) node is fixed at time $t=0$ and carries no momentum. The lower birth node carries weight $B(k)$ with momentum $k$ to be integrated over.}
\label{FeynmanPic}
\end{figure}
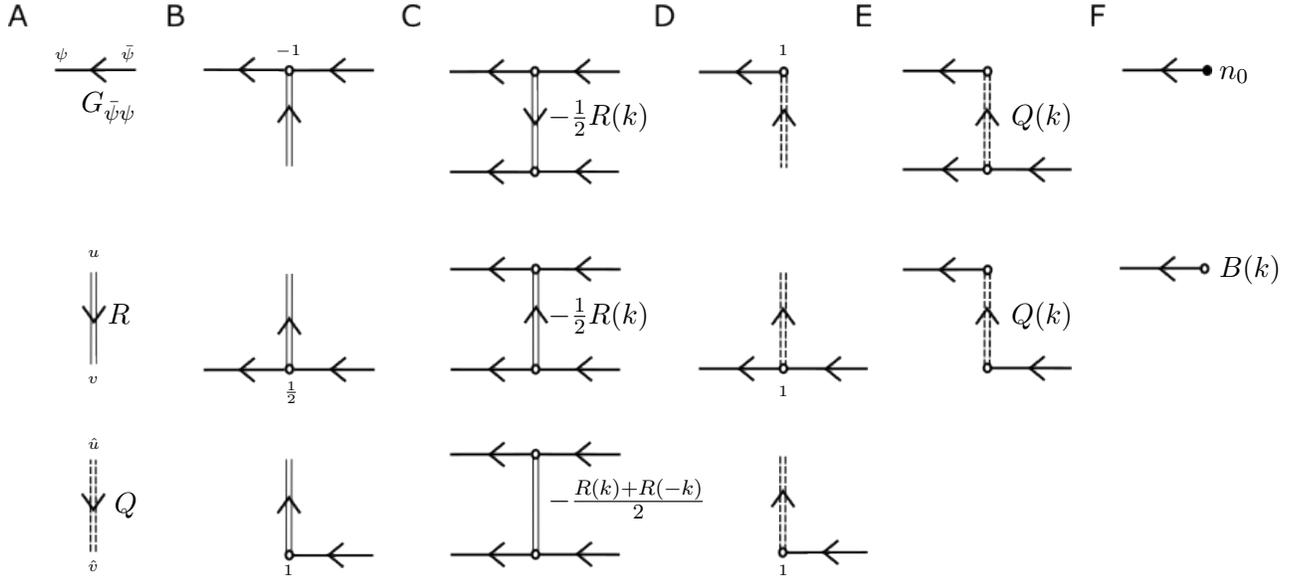

Expansions of the action can be simplified with the aid of a Hubbard-Stratonovich transform. These transforms basically introduce two auxiliary fields into the action, which allows the separation of complicated terms into simpler components \cite{Coleman1988, Hofmann2022, Kopietz1997}. For the  situation in hand, specifically, the following identity is required,
\begin{equation}
\exp\left\{ \int_t\int_p\int_q f_pR_{pq}g_q \right\}
= \mathcal{N}^{-1}\iint \mathcal{D}[u_p]\mathcal{D}[v_q]
\exp \left\{ -\int_t\left[\int_p\int_qu_pR^{-1}_{pq}v_q 
-\int_p u_pg_p-\int_q f_qv_q\right]\right\},
\nonumber
\end{equation}
for some normalisation constant $\mathcal{N}$. The term $R^{-1}$ represents the inverse of function $R$ in the sense that $\int_q R_{pq}R^{-1}_{qr} = \int_q R_{pq}^{-1}R_{qr} = \delta(p-r)$. Note that although $R_{pq}=R(|p-q|)$ is a function representing the interaction profile, the inverse can take the form of an operator or distribution and a bit of care may be needed. Note, however, for the spatially homogeneous functions in Table \ref{Interactions}, this expression becomes a convolution, and so a product in momentum space, meaning $R^{-1}(k)=R(k)^{-1}$.

\begin{figure}[t!]
\centering
\includegraphics[width=0.65\textwidth]{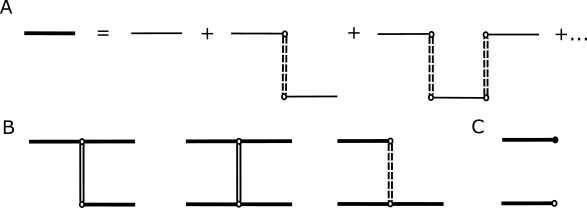}
\caption{A) Dressing of propagator $G_{\bar{\psi}\psi}$ via a Dyson series involving branching propagator $Q$ edges. Feynman graph components are B) final, internal and initiating ladder structures involving annihilation propagator edges $R$ and ladder terminations via a branching propagator edge $Q$, along with C) initiating nodes. Edge directions can be inferred and so are dropped.}
\label{DysonPic}
\end{figure}

Now, there are two terms with double space integrals in Eq. (\ref{ActionFull}) to disentangle, the annihilation term and the branching term, each of which requires a Hubbard-Stratonovich transform. Then if we use $f_p = -\bar{\psi}_p\psi_p$ and $g_q = \psi_q+\frac{1}{2}\bar{\psi}_q\psi_q$ with auxiliary fields $u_p$ and $v_q$ for annihilation, and $\hat{f}_p = \bar{\psi}_p$ and $\hat{g}_q = [\bar{\psi_q}+1]\psi_q$ with auxiliary fields $\hat{u}_p$ and $\hat{v}_q$ for branching, the following action is obtained,
\begin{eqnarray}
S & = & -\int_t\int_p \bar{\psi}_p(\partial_t-D\nabla^2)\psi_p
-\int_t \int_p M_p \bar{\psi}_p \psi_p
+\int_t \int_p B_p \bar{\psi}_p
+n_0\int_p \bar{\psi}_{p}(t=0)
\nonumber \\
&& -\int_t\int_p\int_q u_pR^{-1}_{pq}v_q 
+\int_t\int_p u_p\left(\psi_p+\frac{1}{2}\bar{\psi}_p\psi_p\right)
-\int_t\int_q v_q\bar{\psi}_q\psi_q
\nonumber \\
&& -\int_t\int_p\int_q \hat{u}_pQ^{-1}_{pq}\hat{v}_q
+\int_t\int_p \hat{u}_p\left(\psi_p+\bar{\psi}_p\psi_p\right)
+\int_t\int_q \hat{v}_q\bar{\psi}_q.
\label{ActionHS}
\end{eqnarray}
The first term on each line and the $M_p$ term are quadratic in nature, giving rise to three classes of propagators (one from each line) as seen in Fig. \ref{FeynmanPic}A. Although other quadratic terms (such as $u_p\psi_p$) could be dealt with as propagators, it will prove simple enough to deal with these perturbatively. If we assume a homogeneous death rate, so $M_p=DM$ for some constant $M$, then in momentum space the first propagator is given as follows, where $\theta(t)$ is the standard unit step function,
\begin{align}
G_{\bar{\psi}\psi}(k,\omega) & = \frac{1}{-i\omega+D(k^2+M)},
& G_{\bar{\psi}\psi}(k,t) = \theta(t)e^{-tD(k^2+M)}.
\label{AnnProp}
\end{align}
For the case ($M=0$), this reduces to the standard propagator found in annihilation models \cite{Lee1994}. The other two quadratic terms result in propagators (in momentum space) $R(k)$ and $Q(k)$. The lack of time dependence (for interaction rates) means that the start and end of propagators have the same time point (and so are drawn vertically in Fig. \ref{FeynmanPic}). Note that all interactions we consider are isotropic, meaning in particular that $R(x)=R(-x)$, and so in Fourier space, $R(k)=R(-k)$, meaning the coefficient of the undirected edge in Fig. \ref{FeynmanPic}C can be taken as simply $-R(k)$. The resulting Feynman graph components are summarised in Fig. \ref{DysonPic}B.

Note, firstly, that directions on edges are clear from context and can be dropped. Secondly, the remaining terms, dealt with by perturbative expansion, now only implicate a single position variable. Thirdly, interaction functions and parameters are now only present as propagators, meaning the nodes only have numerical coefficients, except for terminal nodes containing a $B(k)$ term (see Fig. \ref{FeynmanPic}F). Finally, the rules of expansion restricted to annihilation match those found in $\cite{Doi1976, Doi1976b}$, where time ordering (that is, not path integral) formulation of field theory is used. They can also be obtained utilising quantum stochastic process machinery \cite{Greenman2023}.

Now, the lower diagram in Fig. \ref{FeynmanPic}E can be used to convert the propagator $G_{\bar{\psi}\psi}$ into a new propagator $G$ dressed with branching propagators $Q$. This uses the sequence of diagrams in Fig. \ref{DysonPic}A to form a Dyson like sum, where we find, in momentum representation, using Eq. (\ref{AnnProp}),
\begin{eqnarray}
G(k,\omega) & = & G_{\bar{\psi}\psi}(k,\omega)+G_{\bar{\psi}\psi}(k,\omega)Q(k)G_{\bar{\psi}\psi}(k,\omega)
+G_{\bar{\psi}\psi}(k,\omega)Q(k)G_{\bar{\psi}\psi}(k,\omega)Q(k)G_{\bar{\psi}\psi}(k,\omega)+\dots
\nonumber \\
& = & \frac{1}{-i\omega+D(k^2+M)-Q(k)},
\nonumber \\
G(k,t) & = & e^{-t(D(k^2+M)-Q(k))}.
\nonumber
\label{AnnBranchProp}
\end{eqnarray}
Note that the local case $Q(k) = DQ$ recovers the annihilation-branching propagator found in \cite{Cardy1998}. 


\subsection{Mean Field Behaviour}
\label{MFB}

Consider next the mean field behaviour of the two models. These can be obtained either from tree expansions or by assuming that correlation functions resolve into products of single site averages.

\begin{figure}[t!]
\centering
\includegraphics[width=0.65\textwidth]{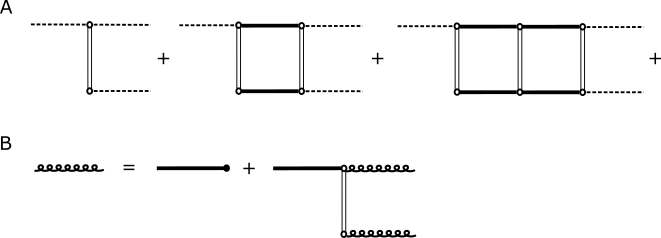}
\caption{Feynman diagrams. A) Diagrams corresponding to the vertex function $\Gamma^{(1,2)}$. Dotted lines indicate truncated propagators. B) Dyson recursion diagrams for tree approximation to mean density.}
\label{FeynmanDiags}
\end{figure}

Firstly, for annihilation Model I ($B_p=M_p=Q_{pq}=0$), consider the tree diagram in Fig. \ref{FeynmanDiags}B. Now, in momentum-time space, $X(k=0,t)$ represents the total density at time $t$. As the external momenta are zero (note Model I has no terminal $B$ nodes, and the terminal $n_0$ nodes also carry no momentum), the internal interaction edge also carries zero momentum and we simply find (letting $X_{cl}$ denote the classical tree approximation to density),
\begin{equation}
X_{cl}(0,t) = n_0G(0,t)+R(0)\int_0^t G(0,t-t')X_{cl}(0,t')^2\dd t'.
\nonumber
\end{equation}
As $G(0,t)=1$, this differentiates to $\frac{\partial X_{cl}}{\partial t}=RX_{cl}^2$, with initial condition $X_{cl}(0,0)=n_0$, resulting in,
\begin{equation}
X_{cl}(0,t) = \frac{n_0}{1+Rn_0t}.
\label{AnnMnF}
\end{equation}

So the end result is just that seen for local interaction. Now the steady state of the annihilation Model I is population extinction (or a lone survivor), and it is the time decay towards this state that is of interest. In particular, it is known \cite{Lee1994} for local interaction and large times (without mean field approximation), that below critical dimension $d_c=2$, the population decay is of order $\mathcal{O}(t^{-d/2})$, whereas above this dimension, it is just of order $\mathcal{O}(t^{-1})$, as seen in the mean field solution in Eq. (\ref{AnnMnF}).

Conversely, for full Model II, the system has a non-zero steady state particle density for sufficiently large birth or branching rates, and it is criticality around the point where non-zero mean population emerges that is of interest (although the less studied decay to this state would also be of interest). Subsequently, it is assumed that sufficient time has passed such that the initial density $n_0$ no longer has influence and can be dropped from the action in Eq. (\ref{ActionFull}). 

If the $n_0$ node is replaced with a $B$ terminal node in Fig. \ref{FeynmanDiags}B, then a similar recursion is found for the tree approximation to mean density for Model II, where, if we assume a homogeneous spontaneous birth rate $B_p = DB$,
\begin{equation}
X_{cl}(k=0,t) = DB\int_0^t G(k=0,t-t') \dd t'
+\int_0^t G(k=0,t-t')RX_{cl}(k=0,t')^2.
\nonumber
\end{equation}
Now if we differentiate this with respect to $t$, after a bit of manipulation we obtain a Ricatti equation $\frac{\partial X_{cl}}{\partial t}=DB+RX_{cl}^2-\tau X_{cl}$, where we have defined $\tau=D(M-Q)$, which leads to steady state,
\begin{equation}
DB = X_{cl}(\tau-RX_{cl}).
\label{RiccSS}
\end{equation}

Thus for the simple case of no spontaneous birth ($B=0$) we find density $X_{cl} = 0$ when $\tau<0$ and $X_{cl} = \frac{\tau}{R}$ when $\tau \ge 0$. 

Thus the same effect results as seen in Model I; the tree structure contains zero (spatial) momentum, interaction terms such as $R(k=0) = R\hat{R}(k=0) = R$ simplify, and the mean field equation observed for local interaction \cite{Janssen1999} results. 

In \cite{Janssen1999}, it is known (without mean field approximation) for local interaction that the position where positive density emerges is shifted from $\tau=0$. The scaling also differs, if $\bar{\tau}$ is the distance above this shifted point, it takes the form $X \propto \bar{\tau}^\beta$ for non-unit constant $\beta = 1-\frac{\epsilon}{6}+\mathcal{O}(\epsilon^2)$ (where $\epsilon = 4-d$).

For both models I and II, seeing the corresponding critical point analysis (without mean field approximation) for non-local modes of interaction are desirable. They are studied in turn over the next two sections.


\section{Renormalisation for Annihilation Model}
\label{AnnRenSec}

Renormalisation methods can be used to deal with divergences seen in perturbative expansions, and also to examine critical exponents in phase transitions  \cite{Bellac1992, Binney1992, Collins1984, Lee1994, Talagrand2022, Tauber2014, Vasil2004}.

In this section, the renormalisation techniques seen in \cite{Lee1994} are first adapted to non-local interaction for annihilation Model I (the addition of branching, spontaneous birth and death in Model II is considered in the Section \ref{SecModII}). Thus $Q_{pq}=M_p=B_p=0$ throughout. In this section, the nature of UV and IR divergences are first highlighted in more detail. It is then shown how solutions to the Callan-Symanzyk equation found in \cite{Lee1994} can be derived by a simple scaling argument without need of the equation itself. This is then used to show that in the asymptotic regime, non-local coupling becomes local and the same asymptotics apply. 

\subsection{Ultra Violet Regulation and Infra Red Divergence}

Given the formalism above, the first thing we note is that the non-local profile acts as a regulator. Consider for example the first and second diagrams for the vertex function $\Gamma^{(1,2)}(k,t)$ in Fig. \ref{FeynmanDiags}A (note that the arms are truncated). In the first diagram, as all external momenta are zero, the annihilation interaction propagator is $R(k=0) = R$ meaning the local interaction term is recovered, that is, in momentum-time space,
\begin{equation}
I_1 = -\int_0^t\dd t' R(0) = -Rt.
\nonumber
\end{equation}

Now, the second term is known to be UV divergent for local interactions, where a primitive dimensional analysis shows it is divergent in dimensions $d \ge 2$. However, for the non-local case, there is an integration over the loop momentum (of dimension $d$) and frequency (of dimension $2$) which is offset by not just two quadratic propagators in the denominator, but also by two interaction terms which may further reduce the divergence depending on their form. The end result is,
\begin{equation}
I_2 = \int_{t>t_2>t_1>0} \fint_k R(-k)G_{\bar{\psi}\psi}(k,t_2-t_1)G_{\bar{\psi}\psi}(-k;t_2-t_1)
R(k).
\nonumber
\end{equation}

If the normal interaction is used, this reduces to the following expressions, where first the $k$ integral is done as standard multivariate normal distribution normalisation, and then the time integrals are calculated directly,
\begin{eqnarray}
I_2 & = & R^2\int_{t>t_2>t_1>0} \fint_k e^{-\frac{k^2}{2\lambda^2}}e^{-2Dk^2(t_2-t_1)}
\nonumber \\
& = & \frac{R^2}{(8\pi)^{\frac{d}{2}}}\left\{ \frac{1}{(2D)^2(2-d)(4-d)}\left[ 
(\lambda^{-2}+4Dt)^{-\frac{d}{2}+2}-\lambda^{d-4}\right] -\frac{\lambda^{d-2}t}{2D(2-d)}\right\}
\nonumber \\
& = & \frac{R^2\lambda^d}{(8\pi)^{\frac{d}{2}}}
\left\{ \frac{t^2}{2!}-\frac{d(2D\lambda^2)t^3}{3!}
+\frac{d(d+2)(2D\lambda^2)^2t^4}{4!}+\dots \right\}.
\nonumber
\end{eqnarray}
Now the second line reduces to $\frac{R^2t^{-\frac{d}{2}+2}}{(8\pi D)^{\frac{d}{2}}(2-d)(4-d)}$ as $\lambda \rightarrow \infty$. This expression can be found in \cite{Tauber2005} for the local interaction. There it is noted that the effective (dimensionless) coupling $I_2/I_1$ is proportional to $t^{-\frac{d}{2}+1}$ which diverges for $d>2$ and small time (this is essentially the UV divergence). For $d<2$ this term also diverges in the (IR) regime of large $t$. Now, for finite $\lambda$, and $t<\frac{1}{4D\lambda^2}$ sufficiently small, a series expansion in $t$ can be found which cancels the removable $d-2$ and $d-4$ singularities and gives a convergent series for any dimension and the UV divergence is naturally regulated. For large $t$ and $d>2$ the ratio $I_2/I_1$ is still well behaved. However, for large $t$ and $d<2$ we find (from the second line above) that we still have an IR divergence to deal with. 

Note that these observations also point towards a cross-over timescale between local and non-local features. More specifically, we have already observed that as the precision parameter $\lambda \rightarrow \infty$, the expression for $I_2$ given by the local model is recovered, as expected. Furthermore, we see that from the condition $t<\frac{1}{4D\lambda^2}$, as $\lambda$ increases towards this limit, the range of $t$ values for which UV regulation exists decreases, and the model naturally becomes more local.

If the screened Poisson profile is used instead, the following integral is obtained for $I_2$. Note that the two interaction propagators (which just become $R^2$ in the local model) each provide an extra quadratic factor in the denominator and primitive divergence estimates suggest convergence for $d<6$. If Schwinger parametrisation \cite{Collins1984, Smirnov2006, Weinzierl2022} is initially used, followed by the $t$ integration (using convolution properties of Laplace transforms), and finally the the momentum integration over $k$ performed, we obtain,
\begin{eqnarray}
I_2 & = & R^2\int_{t>t_2>t_1>0} \fint_k \left(\frac{\lambda^2}{\lambda^2+k^2}\right)^2 e^{-2Dk^2(t_2-t_1)}
= R^2\lambda^4\int_{t>t_2>t_1>0} \fint_k\int_0^\infty \dd \alpha \hh \alpha e^{-\alpha(\lambda^2+k^2)}e^{-2Dk^2(t_2-t_1)}
\nonumber \\
& = & \frac{R^2}{(4\pi)^{\frac{d}{2}}}\int_0^\infty \dd x \hh xe^{-x}\left[ \frac{\left(\frac{x}{\lambda^2}+2Dt\right)^{2-\frac{d}{2}} -\left(\frac{x}{\lambda^2}\right)^{2-\frac{d}{2}}}{D^2(2-d)(4-d)}
-\frac{(\frac{x}{\lambda^2})^{1-\frac{d}{2}}t}{D(2-d)}\right]
\nonumber \\
& = & \frac{R^2\lambda^{d-4}}
{(4\pi)^{\frac{d}{2}}D^2(d-2)(d-4)}
\left\{\lambda^2e^{2Dt\lambda^2}
\left[\Gamma\left(4-\frac{d}{2},2Dt\lambda^2\right)
-2Dt\lambda^2\Gamma\left(3-\frac{d}{2},2Dt\lambda^2\right)\right] \right.
\nonumber\\ 
&& \hspace{8cm}
-\left.\left[\Gamma\left(4-\frac{d}{2}\right)
+(4-d)Dt\lambda^2\Gamma\left(3-\frac{d}{2}\right)\right]\right\}.
\nonumber
\end{eqnarray}
In the second line above the transform $y=\frac{x}{\lambda^2}+2Dt$ is used in the first term and then upper incomplete gamma functions can be used. Note both factors $d-2$ and $d-4$ are removable singularities, which can be seen by L'Hopital's rule upon the second line. For finite $t$ the incomplete upper gamma functions are entire and contain no divergences. The only divergences arise from the final two terms for dimensions $d=6,8,10,\dots$ and the term is otherwise regulated. As $t$ approaches zero, the incomplete gamma functions become complete (with poles), also presenting divergences at these dimensions. This matches the primitive dimension of divergence. Thus the UV regulation is found to be weaker than for the normal interaction, which has an exponentially decaying tail and the integral is convergent in any dimension. For $d<2$ we again find IR behaviour for large $t$. 

The cross-over timescale between local and non-local features can also be observed for the screened Poisson form of interaction. Firstly, from the second line in the equation above, we again see that in the limit of $\lambda\rightarrow\infty$ the term $\frac{R^2t^{-\frac{d}{2}+2}}{(8\pi D)^{\frac{d}{2}}(2-d)(4-d)}$ for the local model is recovered. Furthermore, from the same line we find expansion $\frac{1}{\varepsilon}[(\frac{x}{\lambda^2}+2Dt)^{\frac{\varepsilon}{2}}-(\frac{x}{\lambda^2})^{\frac{\varepsilon}{2}}] = \log\sqrt{1+\frac{2Dt\lambda^2}{x}}+\mathcal{O}(\varepsilon)$ (where $\varepsilon = 4-d$), demonstrating the removable singularity at dimension $d=4$ for finite $\lambda$. We see that only the leading term remains at this dimension, which diverges as $\lambda\rightarrow\infty$, as would be expected. Note that if $\lambda$ is increased, $t$ must decrease to maintain a fixed value for this term. This implies that as $\lambda$ is increased and the model becomes more localised, the range of $t$ values exhibiting effective UV regulation is reduced. 

For the spherical interaction, a precise calculation is somewhat intractable, although we note the following. For general interaction profile $R(k)=R\hat{R}(k)$ we find that,
\begin{eqnarray}
I_2 & = & R^2\int_{t>t_2>t_1>0}\fint_k \hat{R}(k)^2 e^{-2Dk^2(t_2-t_1)} = 
\frac{2R^2}{(4\pi)^{\frac{d}{2}}\Gamma(\frac{d}{2})}\int_0^\infty \dd k \hh k^{d-1}\hat{R}(k)^2
\left[ \frac{t}{2Dk^2} - \frac{1}{4D^2k^4}
+\frac{e^{-2Dk^2t}}{4D^2k^4}\right]
\nonumber \\
& = & \frac{R^2}{(8\pi D)^{\frac{d}{2}}\Gamma(\frac{d}{2})}t^{2-\frac{d}{2}}
\int_0^\infty \dd x \hh x^{\frac{d}{2}-1}\hat{R}\left( \sqrt{\frac{x}{2Dt}} \right)^2 \frac{x-1+e^{-x}}{x^2}.
\nonumber
\end{eqnarray}
Now, any form of asymptotic decay of the function $\hat{R}(\sqrt{\frac{x}{2Dt}})$ will render the integral convergent at critical dimension $d_c=2$. Any positive power law for $t$ from this term will also push the divergence of $I_2/I_1$ as $t\rightarrow 0$ to dimensions above $d_c$ and UV divergence will be regulated. 

Then, for the spherical interaction, we have the asymptotic form given by $R(k)\propto (\frac{k}{\lambda})^{-\frac{d}{2}}J_{\frac{d}{2}}(\frac{k}{\lambda})\sim (\frac{k}{\lambda})^{-\frac{d}{2}}\sqrt{\frac{2\lambda}{\pi k}}\cos(\frac{k}{\lambda}-(d+1)\frac{\pi}{4})$. Note that the higher the dimension, the stronger the asymptotic decay. We subsequently find that integral $I_2$ is finite in any dimension and $I_2/I_1\sim\mathcal{O}(\lambda^{d+1} t^{\frac{3}{2}})$ meaning UV divergence is regulated in any dimension for small enough $t$, although we again will have IR behaviour for asymptotic large $t$. 

Consider the cross-over time scales between local and non-local behaviour for this spherical interaction profile. The $\mathcal{O}(\lambda^{d+1} t^{\frac{3}{2}})$ term means that as $\lambda$ increases and the model becomes increasingly local (and this term more divergent), maintaining the UV regulation is restricted to smaller values of $t$. An exact expression for the calculation of $I_2$ would make the cross-over time scale here a bit more transparent. 

In summary, we have shown that when the precision parameter is finite, various forms of non-local interaction act as a natural form of UV regulation (in the renormalisation sense). This suggests control of UV divergences is possible provided the interaction function tails off sufficiently quickly in momentum representation. We have also seen that as the precision parameter $\lambda$ for any of these non-local models gets larger, the regulation weakens as the interaction tends toward local behaviour and UV divergences start to emerge. Even for finite precision though, IR behaviour is still present for large $t$, for which we turn to rescaling methods that underlie the renormalisation group. 


\subsection{Rescaling}

Here it is shown that the renormalisation group calculation in \cite{Lee1994} can also be found via a simple rescaling of the action in Eq. (\ref{ActionFull}). Specifically, if we perform the following space-time-field transforms,
\begin{align}
p' & = \gamma p,
& \psi' & = \alpha \psi, \nonumber \\
t' & = \eta t,
& \bar{\psi}' & = \beta \bar{\psi},
\label{STFRescale}
\end{align}
and require that the action structure in Eq. (\ref{ActionFull}) is unchanged (the top line of the equation, corresponding to Model I), albeit with new parameters. More precisely, the first term $\int_t\int_p \bar{\psi}_p(\partial_t) \psi_p$ in Eq. (\ref{ActionFull}) rescales to $\gamma^{-d}\alpha^{-1}\beta^{-1}\int_{t'}\int_{p'} {\bar{\psi}_{p'}}'(\partial_{t'}) {\psi_{p'}}'$, which in turn leads to condition $\gamma^d\alpha\beta=1$. Similarly, the second term $\int_t\int_p \bar{\psi}_p(D\nabla^2)\psi_p$ transforms to $\gamma^{2-d}\eta^{-1}\alpha^{-1}\beta^{-1}\int_{t'}\int_{p'} {\bar{\psi}_{p'}}'(D\nabla^2){\psi_{p'}}'$ and the structure is preserved provided we define an updated diffusion as $D' = \eta^{-1}\gamma^2D$. For the term involving interaction $R_{pq}$, we note from Table \ref{Interactions} that this can be quite broadly assumed to take the form $R\hat{R}(p-q,\lambda)$, where $\hat{R}$ is a normalised function of space (so Fourier transform $\hat{R}(k=0,\lambda)=1$), spatial dependencies are homogeneous, and $\lambda$ is a form of precision. Such functions have scaling properties,
\begin{align}
\hat{R}(ap,\lambda) & = a^{-d}\hat{R}(p,a\lambda), 
& \hat{R}(a^{-1}k,\lambda) & = \hat{R}(k,a\lambda). 
\label{IntSc}
\end{align}
Now if these properties are used in conjunction with the scaling in Eq.s (\ref{STFRescale}), then the interaction term of Eq. (\ref{ActionFull}) is conserved if we define new interaction $R'=\alpha^{-1}\eta^{-1}R$ and precision $\lambda' = \gamma\lambda$. Note also, this term contains cubic and quartic interactions (the quartic having an extra $\bar{\psi}$ field term), and so can only conserve its form if $\beta=1$, which leads to $\alpha = \gamma^{-d}$.

Together then, the parameters must update as,
\begin{align}
D' & = \tau^{-1}\gamma^2D,
& \lambda' & = \gamma^{-1}\lambda, \nonumber \\
R' & = \gamma^{d}\eta^{-1}R ,
& n_0' & = \gamma^{-d}n_0.
\label{ParamRescale}
\end{align}
The condition on the diffusion parameter sets how time and space scale together. The natural choice is to fix diffusion $\eta = \gamma^2$, which essentially puts time squared and space on the same dimensional footing, which is what standard dimensional analysis does. This has the important consequence that $R' = \gamma^{d-2}R$ and we see the critical scaling dimension $d_c=2$ emerge. 

Note that the space-time-field rescaling results in a non-unit Jacobian to  path integral expressions such as Eq. (\ref{MeanDen}), but this same factor appears in the path integral normalisation constant $\mathcal{N}$ and thus cancels. Given the path integral definition of the mean density in Eq. (\ref{MeanDen}), the following rescaling law is obtained (where the invariant diffusion constant $D=D'$ is dropped from the notation),
\begin{equation}
X(p,t;R,\lambda,n_0) = \gamma^d
X(p\gamma,t\gamma^{2};R\gamma^{d-2},\lambda\gamma^{-1}, n_0\gamma^{-d}).
\nonumber
\end{equation}

Without renormalisation, this form of rescaling is not too involved. For example, it is a simple exercise to show  the mean field solution in Eq. (\ref{AnnMnF}) satisfies this property. However, once renormalization is set at some scale, establishing this form of invariance is less straightforward. This expression is also somewhat akin to the solution to the Callan-Symanzyk equations seen in \cite{Lee1994}. The main differences are firstly, there is rescaling of the extra parameter $\lambda$, and secondly, that it is in terms of the original interaction parameter $R$ rather than a renormalised version. However, we can recover the Callan-Symanzyk solution via the simple scaling above (without solving or needing the Callan-Symanzyk PDE) as follows.

In the case of local interactions, the following vertex function is obtained in momentum-Laplace space (see \cite{Lee1994} for details),
\begin{equation}
\Gamma^{(1,2)}(k,s) = \frac{R}{1+R(g^{*})^{-1}(s+\frac{1}{2}k^2)^{\frac{d}{2}-1}}.
\label{Vertex}
\end{equation}
The constant $g^* =\frac{(4\pi)^{1-\frac{\epsilon}{2}}} {\Gamma(\frac{\epsilon}{2})}$ is of order $\mathcal{O}(\epsilon)$. This is just the expression given in \cite{Lee1994} (which considers the more general (but local) $K$ particle process $KA \rightarrow \phi$) restricted to the case $K=2$. Next we select a fixed momentum scale $\kappa$ by defining $\kappa^2 = s+p^2/2$ for some $s$ and $p$. In \cite{Lee1994} and other applications it is convenient to pick zero momentum $p=0$ (and so $s=\kappa^2$), although the specific choice is immaterial and somewhat arbitrary. This is used to define a renormalised parameter $R_r$,
\begin{equation}
R_r = \frac{R}{1+R(g^*)^{-1}\kappa^{d-2}}.
\nonumber
\end{equation}
Note that $R_r$ is simply the value of a vertex function at this stage. Next the momentum scale $\kappa$ is used to define dimensionless parameters, by defining $g=R\kappa^{d-2}$ and $g_r=R_r\kappa^{d-2}$ to give,
\begin{equation}
g_r = \frac{g}{1+g(g^*)^{-1}}.
\label{gr_val}
\end{equation}
Now, if a new set of parameters is chosen in accordance to the rescaling of Eq. (\ref{ParamRescale}) (designated with an apostrophe), there is a corresponding renormalised parameter $g_r'$ (which we define with the same momentum scale $\kappa$). Then from $R' = \gamma^{d-2}R$ and so $g' = \gamma^{d-2}g$,
\begin{equation}
g_r' = \frac{g\gamma^{d-2}}{1+g\gamma^{d-2}(g^*)^{-1}}.
\label{gr_dash_val}
\end{equation}
It is now possible to eliminate $g$ from Eq.s (\ref{gr_val}) and (\ref{gr_dash_val}) and after a little algebra obtain (note $\gamma^2 = \frac{t'}{t}$),
\begin{equation}
\frac{g^*}{g_r'}-1 = \left(\frac{t}{t'}\right)^{d-2}
\left( \frac{g^*}{g_r}-1 \right).
\nonumber
\end{equation}
This is precisely the form seen in \cite{Lee1994} as part of the Callan-Symanzik solution. In particular, for $d<d_c=2$ we find that as $t \rightarrow \infty$ (so $\gamma\rightarrow 0$), the limit $g_r' \rightarrow g^*$ arises, whereas for $d> d_c$ the limit $g_r' \rightarrow 0$ occurs. These are precisely the fixed points of the Callan-Symanzyk equations in \cite{Lee1994}. Note also that the relationship between $g_r$ and $g_r'$ as we shift between $t$ and $t'$ is independent of the choice $\kappa$, which does nothing other than set the scale and initialise the value $g_r$. Finally, we note that if the $R$ dependency is replaced with the dimensionless parameter $g_r$ (and we write $g_r'=g_r(\gamma)$), the rescaling law in \cite{Lee1994} is found, that is, the full solution to the Callan-Symanzyk equation for the density;
\begin{equation}
X(p,t;g_r,\lambda,n_0) = \gamma^{d}
X(p\gamma,t\gamma^{2};g_r(\gamma),\lambda\gamma^{-1}, n_0\gamma^{-d}).
\label{XCS}
\end{equation}
Conversely, this is telling us that running along the characteristics of solutions to Callan-Symanzik equations is equivalent to the action preserving rescaling of Eq. (\ref{ParamRescale}).

Note that from $\gamma^2 = \frac{t'}{t}$ we find that for large time, the precision parameter, scaling as $\lambda\gamma^{-1}$, becomes large, and the interaction profiles become local. This ultimately explains why the asymptotics of non-local systems are the same as local systems, although the cross-over time scale of such behaviours is a natural question to consider, as briefly explored in the previous section. 

In order to fully deal with IR divergences, the response functional is also required, which dresses the main propagator with all possible tree attachments. This can be obtained in full generality for non-local interactions.


\subsection{The Response Functional}

\begin{figure}[t!]
\setlength{\unitlength}{0.1\textwidth}
\begin{picture}(9,3.5)
\put(0.00,0.00){\includegraphics[width=16cm]{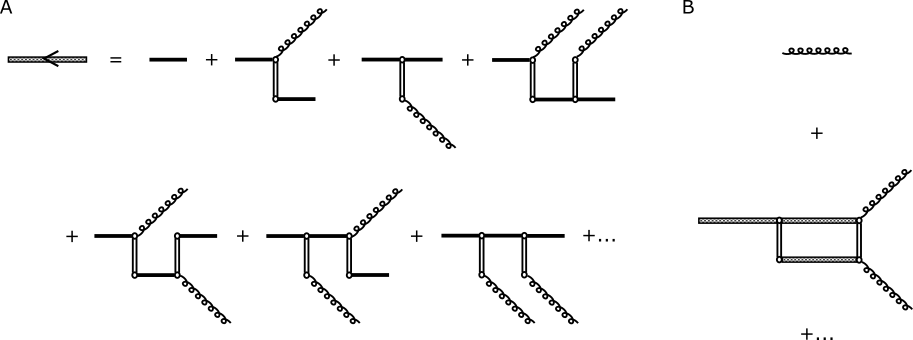}}
\put(0.05,2.92){{\tiny $t_2$}}
\put(0.75,2.92){{\tiny $t_1$}}
\put(0.40,2.62){{\tiny $k$}}
\put(2.35,2.55){{\tiny $R(k)$}}
\put(3.62,2.55){{\tiny $R(0)$}}
\put(4.92,2.55){{\tiny $R(k)$}}
\put(5.37,2.55){{\tiny $R(0)$}}
\put(1.37,0.77){{\tiny $R(k)$}}
\put(0.93,0.77){{\tiny $R(k)$}}
\put(2.67,0.77){{\tiny $R(0)$}}
\put(3.09,0.77){{\tiny $R(k)$}}
\put(4.42,0.77){{\tiny $R(0)$}}
\put(4.85,0.77){{\tiny $R(0)$}}
\end{picture}
\centering
\caption{A) Feynman diagrams for the response functional up to second order. Time runs from right to left. The texture filled edge represents the response functional $G_{cl}(k;t_1,t_2)$ from time $t_1$ to $t_2$ for momentum $k$. The spiral edges indicate tree structures which all extend from time $0$ and carry zero momentum. Black line edges are propagators and all carry momentum $k$. Double edges are interaction propagators, with momenta stated (either zero or $k$). B) The first two diagrams for the mean density using response functionals.}
\label{ResponsePic}
\end{figure}
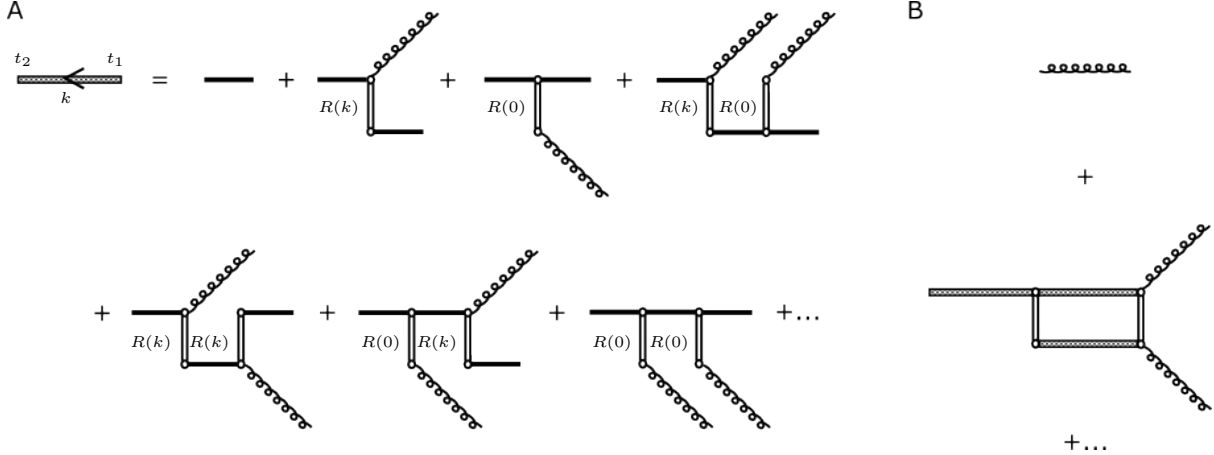

The response functional is the standard propagator running from time $t_1$ to $t_2$ with any number of tree branches attached, where each tree runs from time $0$ to the time of attachment somewhere in $[t_1,t_2]$. The first few corresponding diagrams can be seen in Fig. \ref{ResponsePic} (represented as spirals). The external momenta running along attached trees are taken to be zero (this essentially means the initial densities $n_0$ at the end of trees are spatially uniform). The momenta running along the response functional, $k$ (which may be part of a loop in a larger diagram, for example), can be non-zero. Note that each each time point of attachment, we have two possible nodes to attach the tree  branch to. This results in two possible choices for the interaction propagator, either $R(0)$ or $R(k)$, leading to the expression,
\begin{align}
G_{cl}(k,t_2,t_1) & = e^{-Dk^2(t_2-t_1)}\exp\left\{ -(R(0)+R(k))\int_{t_1}^{t_2} \dd t \hh X_{cl}(k=0,t)\right\}
\nonumber \\
& = e^{-Dk^2(t_2-t_1)}
\left( \frac{1+Rn_0t_1}{1+Rn_0t_2} \right)^\frac{R(0)+R(k)}{R}
= e^{-Dk^2(t_2-t_1)}\left( \frac{1+Rn_0t_1}{1+Rn_0t_2} \right)^{1+\hat{R}(k)}. 
\nonumber
\end{align}
Note the following. Firstly, the exponent in the final term is independent of $R$, and just reduces to the standard value $2$ for local interaction $R(k)=R$. Secondly, from Eq. (\ref{AnnMnF}), we have seen that the classical (tree) density $X_{cl}(t)$ takes the form given by a local interaction irrespective of how non-local the actual interaction is. Finally, note from Fig. \ref{DysonPic} that if branching, birth and death is allowed in addition to annihilation, the diagrams in Fig. \ref{ResponsePic} will not change, meaning the response functional will have similar structure (the first exponential would be modified to $\exp\left\{-(t_2-t_1)[D(k^2+M)-Q(k)]\right\}$ and $X_{cl}$ would be the solution to the Ricatti equation above Eq. (\ref{RiccSS})). As only steady state will be explored for Model II, the response functional will not be needed there and is not explored further.

\subsection{Critical Behaviour}

It is now possible to put things together and examine the asymptotic long time behaviour of the density. Let us summarise the pieces. Now, the particle density $X(t;g,\lambda,n_0)$ (the $p$ dependency is dropped due to spatial uniformity) can be viewed as a perturbative expansion in terms of the un-renormalised interaction parameter $g$, a series which contains UV divergences. Next renormalisation via Eq. (\ref{gr_val}) converts variable $g$ to the renormalised parameter $g_R$. Re-expressing the density $X(t;g_R,\lambda,n_0)$ in terms of $g_R$ then results in a series free of UV divergence. However, for large times (which we want), IR divergences remain, whereas for sufficiently small $t$, this series is convergent. The last part of the process is then to use the renormalisation group (or equivalently, the scaling effects described above) to link the IR divergent domain to a convergent one. 

Now, we can write Eq. (\ref{XCS}) as,
\begin{equation}
X(t;g_r,\lambda,n_0) = \gamma^{d}
(X^{(0)}(t(\gamma);g_R(\gamma),\lambda(\gamma),n_0(\gamma))+X^{(1)}(t(\gamma);g_R(\gamma),\lambda(\gamma),n_0(\gamma))+\cdots),
\nonumber
\end{equation}
where $X^{(0)}$ is the loop-less (tree) contribution to the density, $X^{(1)}$ is the single loop contribution, and so on. Note that the left hand side is asymptotic in $t$ with non-local interaction, as precision $\lambda<\infty$ is finite. On the right hand side, we have a small value $\gamma = \sqrt{\frac{t(\gamma)}{t}}$ to keep $t(\gamma)$ fixed and finite as $t$ gets large. From Eq. (\ref{gr_dash_val}) this means that $g_r(\gamma)\rightarrow g^*$ approaches the fixed point, the precision $\lambda(\gamma) \rightarrow \infty$ gets large, suggesting all terms on the right hand side become identical to those seen in local interaction, returning the terms seen in \cite{Lee1994}, and implying the same universal behaviour is observed (to leading order, at least).

It is usually argued that above the critical dimension, the higher dimensionality means statistical fluctuations are negligible and only the uncorrelated residual mean field behaviour is relevant, where particles interact at the average rate of all others, as seen in tree level expansions. Although reasonable and intuitive, these descriptions are qualitative and when supra-critical effects are discussed in technical detail, they tend to revolve around limiting behaviour using renormalisation group ideas rather than statistical fluctuation negligibility \cite{Binney1992,Hofmann2022}.

Consider the cases either side of the critical dimension in a bit more detail. Now the tree expansion from Eq. (\ref{AnnMnF}) gives,
\begin{equation}
\gamma^dX^{(0)}(t(\gamma);g(\gamma),\lambda(\gamma),n_0(\gamma)) = 
\gamma^{d}\frac{n_0(\gamma)}{1+n_0(\gamma)g(\gamma)\kappa^{d-2}t(\gamma)}
=\gamma^{d}\frac{n_0(\gamma)}{1+n_0(\gamma)\left(\frac{1}{g_R(\gamma)}-\frac{1}{g^*}\right)^{-1}\kappa^{d-2}t(\gamma)}.
\nonumber
\end{equation}
As noted previously, at the tree level, everything is equivalent to local interactions, and we subsequently lack any $\lambda$ dependency. Now for small $\gamma$, $n_0(\gamma)=n_0\gamma^{-d}$ is large and we find,
\begin{eqnarray}
\gamma^dX^{(0)}(t(\gamma);g_R(\gamma),\lambda(\gamma),n_0(\gamma)) & \simeq & 
\left(\frac{1}{g_R(\gamma)}-\frac{1}{g^*}\right)\frac{\gamma^d}{\kappa^{d-2}t(\gamma)} =
\left(\frac{1}{g_R(\gamma)}-\frac{1}{g^*}\right)\frac{t(\gamma)^{\frac{d}{2}-1}}{\kappa^{d-2}t^{\frac{\gamma}{2}}} 
\nonumber \\
& = & \frac{1}{t^{\frac{d}{2}}}\left(\frac{1}{g_R(\gamma)}-\frac{1}{g^*}\right),
\nonumber
\end{eqnarray}
where in the last step $t(\gamma)$ is fixed to the value $\kappa^2$. Now this expression is accurate to order $\mathcal{O}(\frac{1}{g_R(\gamma)})$. The term of constant order $\mathcal{O}(g_R(\gamma)^0)$ is divergent as $\epsilon\rightarrow 0$ due to $g^*=\mathcal{O}(\epsilon)$. However, single loop terms also give a contribution at this order.

The next term, then, $X^{(1)}$ arises from the single loop diagram in Fig. \ref{ResponsePic}B where we find that,
\begin{align}
X^{(1)}(t;R,\lambda,n_0) = & \int_{0<t_1<t_2<t}\fint \dd k
G_{cl}(0,t-t_2)R(k,\lambda)G_{cl}(k;t_2-t_1)^2R(k,\lambda)X_{cl}(t_1)^2
\nonumber \\
= & \int_{0<t_1<t_2<t}\fint \dd k
\left(\frac{1+Rt_2n_0}{1+Rtn_0}\right)^{1+\hat{R}(k,\lambda)}\hat{R}(k,\lambda)\cdot
\nonumber\\
& \hspace{2.5cm}
e^{-Dk^2(t_2-t_1)}\left(\frac{1+Rt_1n_0}{1+Rt_2n_0}\right)^{2+2\hat{R}(k,\lambda)}\hat{R}(k,\lambda)
\left(\frac{n_0R}{1+n_0Rt_1}\right)^2.
\nonumber
\end{align}
Now we have $g=R\kappa^{d-2}$ and scalings $g(\gamma)=g \gamma^{d-2}$, $n_0(\gamma) = n_0\gamma^{-d}$, meaning $Rn_0$ scales to $Rn_0\gamma^{-2}$. Thus as $\gamma\rightarrow 0$ these terms become large, and noting also that $\hat{R}(k,\lambda) \rightarrow 1+\mathcal{O}(\gamma^2)$, under rescaling the $X^{(1)}$ term just reduces to the calculation in \cite{Lee1994},
\begin{eqnarray}
\gamma^d X^{(1)}(t(\gamma);g_r(\gamma),\lambda(\gamma),n_0(\gamma))
& \rightarrow & \frac{\gamma^d}{t(\gamma)^2}\int_{0<t_1<t_2<t(\gamma)}\fint \dd k
\left(\frac{t_1}{t_2}\right)^2e^{-Dk^2(t_2-t_1)}
\nonumber \\
& = & \gamma^d t(\gamma)^{-\frac{d}{2}}\left(\frac{1}{g^*}-\frac{2C+5}{16\pi}+\dots\right)
= t^{-\frac{d}{2}}\left(\frac{1}{g^*}-\frac{2C+5}{16\pi}+\dots\right),
\nonumber
\end{eqnarray}
where $C$ is Euler's constant. Then the divergent term $\frac{1}{g^*}$ cancels when $X^{(0)}$ and $X^{(1)}$ are added and renormalisation has done its job.

Now for the case $\epsilon>0$ below the critical dimension, we find $g_R(\gamma)\rightarrow g^*$ as $\gamma\rightarrow 0$, and the same asymptotics as \cite{Lee1994} result, with density decaying as $X(t)=\mathcal{O}(t^{-\frac{d}{2}})$. 

Conversely, above the critical dimension ($\epsilon<0$), from Eq. (\ref{gr_dash_val}) a small $\gamma$ means that $g_r(\gamma) \simeq g\gamma^{d-2}$. This means the leading order term is,
\begin{equation}
\frac{1}{t^{\frac{d}{2}}g_R(\gamma)} = 
\frac{\gamma^{2-d}}{t^{\frac{d}{2}}g} = 
\left(\frac{t(\gamma)}{t}\right)^{1-\frac{d}{2}}\frac{1}{t^{\frac{d}{2}}g} = \frac{1}{t}\frac{1}{g\kappa^{2-d}} = \frac{1}{Rt}.
\nonumber
\end{equation}
Thus the tree level asymptotics $\mathcal{O}(t^{-1})$ is recovered as expected. Note that the remaining higher order terms $\mathcal{O}(t^{-\frac{d}{2}})$ will decay faster for $d>d_c=2$. 


\section{Annihilating Walks with Birth and Death}
\label{SecModII}

The annihilation Model I in Section \ref{AnnRenSec} only utilises some aspects of renormalization. Specifically, vertex functions can be summed in entirety and propagators require no normalisation (or more specifically, the vertex function $\Gamma^{(1,1)}$ is just a single diagram of the original propagator, no other diagrams start and finish with a single edge). As such it only uses additive renormalisation, no multiplicative renormalisation is needed. Next then, we consider full Model II (see Eq. (\ref{ModDes})), which will requires additive and multiplicative methods of renormalisation. For this model, death, spontaneous birth, and branching are added to the annihilation process. Then any particle at position $p$ can spawn a new one at position $q$ in some non-local, isotropic fashion with rate $Q(|p-q|)$. Additionally, a death rate $M_p$ and spontaneous creation rate $B_p$ is also supposed. This is a non local version of annihilating and branching random walks \cite{Cardy1998}. The Liouvillian for Model II was given in Eq. (\ref{LiouvFull}), resulting in the action of Eq. (\ref{ActionFull}).

Now, when Model I was analysed, a space-time-parameter transformation was used to effectively solve the Callan-Symanzik equation. The same technique can be applied to Model II. However, we need to broaden the class of actions slightly. To this end, the action from Eq. (\ref{ActionFull}) for the full model is rewritten as follows,
\begin{eqnarray}
S & = & \int_t \int_p \bar{\psi}_p(\partial t - D\nabla^2)\psi_p
-\frac{1}{2} \int_t \int_p \int_q R^{(4)}_{pq}\bar{\psi}_p\psi_p\bar{\psi}_q\psi_q
-\int_t \int_p \int_q R^{(3)}_{pq}\bar{\psi}_p \psi_p \psi_q
\nonumber \\
&& +\int_t \int_p \int_q Q^{(3)}_{pq}\bar{\psi}_p \bar{\psi}_q \psi_q
+\int_t \int_p \int_q Q^{(2)}_{pq}\bar{\psi}_p \psi_q
+\int_t \int_p M_p\bar{\psi}_p \psi_p
+\int_t \int_p B_p\bar{\psi}_p.
\nonumber
\end{eqnarray}
Note the initial density term $n_0\int_p \bar{\psi}_p(t=0)$ is ignored in the action. This is equivalent to assuming the system has reached long term steady state, and the initial density $n_0$ no longer has any influence over current behaviour.

Note, from Eq. (\ref{ActionFull}), we initially have equality between the terms $R^{(4)}_{pq}=R^{(3)}_{pq}$ as well as $Q^{(3)}_{pq}=Q^{(2)}_{pq}$, which are the non-local annihilation and birth rates. The superscripts indicate the order of the associated field terms ($R^{(4)}_{pq}$ is quartic for example). The reason that they have distinct superscript labels is that it will be seen that under the space-time-field scaling (identical to that seen in the previous section for pure annihilation), although the general structure is preserved, terms such as $R^{(4)}_{pq}$ and $R^{(3)}_{pq}$ will no longer be required to be equal. The following initial forms of interaction (prior to scaling) are assumed. Specifically, spatial homogeneity is assumed and non-local interactions are isotropic functions of distance separation:
\begin{align}
R^{(3)}_{pq} & = DR^{(3)}\hat{R}(p-q,\lambda_r),
& Q^{(2)}_{pq} & = DQ^{(2)}\hat{Q}(p-q,\lambda_q), 
& M_p & = DM, \nonumber \\
R^{(4)}_{pq} & = DR^{(4)}\hat{R}(p-q,\lambda_r),
& Q^{(3)}_{pq} & = DQ^{(3)}\hat{Q}(p-q,\lambda_q),
& B_p & = DB. \nonumber
\end{align}
Here $\hat{R}$ and $\hat{Q}$ are normalised distributions, so the Fourier transforms satisfy $\hat{R}(k=0,\lambda_r)=\hat{Q}(k=0,\lambda_q)=1$. Note in particular that both of these functions have scaling properties seen in Eq. (\ref{IntSc}).


\subsection{Rescaling}

Next, then, the same rescaling of space-time and fields is implemented as seen for the annihilation model. Specifically,
\begin{align}
p' & = \gamma p,
& \psi' & = \alpha \psi, \nonumber \\
t' & = \eta t,
& \bar{\psi}' & = \beta\bar{\psi}.
\label{Rescaling2}
\end{align}

Now, it is required that the action of Eq. (\ref{ActionFull}) is structurally invariant under this transformation. For the terms occurring in the annihilation model, the same restrictions result, that is; $\gamma^d \alpha\beta=1$ and $D' = \gamma^2\eta^{-1}D$. Continuing this procedure through the remaining terms in the same manner as the previous section (including some manipulation with Eq. (\ref{IntSc})), we find,
\begin{align}
\alpha\beta\gamma^d & = 1, & {Q^{(2)}}' & = \gamma^{-2}Q^{(2)}, 
&& \nonumber \\
D' & = \gamma^{2}\eta^{-1} D, & {Q^{(3)}}' & = \gamma^{-2}\beta^{-1}Q^{(3)},
&\lambda_{q'}' & = \gamma^{-1}\lambda_q,\nonumber \\
M' & = \gamma^{-2}M, & {R^{(3)}}' & = \gamma^{-2}\alpha^{-1} R^{(3)}, 
&\lambda_{r'}' & = \gamma^{-1}\lambda_r,\nonumber \\
B', & = \gamma^{-2-d}\beta^{-1}B & {R^{(4)}}' & = \gamma^{d-2} R^{(4)}.
&&
\label{ParamShift2}
\end{align}
Some comments are worth making. Firstly, the scaling factor $\eta$ only appears in the $D$ transform. This is in part due to the fact that all terms (such as $M_p=DM$, for example) have a $D$ factor (analogous to the choices made in \cite{Janssen1999}). Again, like Model I, the natural choice to make is $\eta = \gamma^2$ which fixes diffusion, and renders time to have the same dimension as momentum squared. Secondly, the field renormalisation factors $\alpha$ and $\beta$ give some flexibility, although the quadratic and quartic terms are functions of $\alpha\beta=\gamma^{-d}$ and immune to any choice. Thirdly, notice that as $\gamma$ gets small, the variable ${R^{(4)}}'$ will become small and irrelevant provided $d>2$. Thus it will not be a relevant variable near a critical dimension $d_c=4$, although, for low dimensions $d<2$, its effects would have to be considered. Following \cite{Janssen2005,Tauber2005}, it is natural to pick $\alpha$ and $\beta$ such that both cubic terms have the same coefficient and so scaling behaviour. Specifically, they are set as:
\begin{align*}
\alpha & = \gamma^{-\frac{d}{2}}\sqrt{\frac{R^{(3)}}{Q^{(3)}}},
&
\beta & = \gamma^{-\frac{d}{2}}\sqrt{\frac{Q^{(3)}}{R^{(3)}}},
\nonumber
\end{align*}
which results in the scaled coefficients,
\begin{equation}
{R^{(3)}}' = {Q^{(3)}}' = \gamma^{\frac{d}{2}-2}\sqrt{R^{(3)}Q^{(3)}}.
\label{RScale}
\end{equation}
Then we can define new variables and get the following scaling behaviour,
\begin{align}
g & = \sqrt{R^{(3)}Q^{(3)}},
& g' & = \gamma^{\frac{d-4}{2}}g,
\nonumber \\
b & = \sqrt{\frac{R^{(3)}}{Q^{(3)}}}B,
& b' & = \gamma^{-2-\frac{d}{2}}b.
\nonumber
\end{align}
Note the scaling for $g$ flips at the critical dimension $d_c=4$.

Then updating the action, we find,
\begin{eqnarray}
S & = & \int_t \int_p \bar{\psi}_p(\partial t - D\nabla^2)\psi_p
-\frac{1}{2} \int_t \int_p \int_q R\hat{R}_{pq}\bar{\psi}_p\psi_p\bar{\psi}_q\psi_q
-\int_t \int_p \int_q g\hat{R}_{pq}\bar{\psi}_p \psi_p \psi_q
\nonumber \\
&& +\int_t \int_p \int_q g\hat{Q}_{pq}\bar{\psi}_p \bar{\psi}_q \psi_q
+\int_t \int_p \int_q Q\hat{Q}_{pq}\bar{\psi}_p \psi_q
+\int_t \int_p M_p\bar{\psi}_p \psi_p
+\int_t \int_p b\bar{\psi}_p,
\nonumber
\end{eqnarray}
where we define $R = R^{(4)}$ and $Q = Q^{(2)}$.

Note from Eq. (\ref{RScale}) that as $\gamma \rightarrow 0$ the scaled precisions $\lambda'$ diverge, the interaction profiles become local, and UV divergences start to materialise. 

Although the action structure is preserved, some terms are scaled, in particular the mean density, where it is found that,
\begin{equation}
X(p,t;R,g,Q,M,b,\lambda) = \gamma^{\frac{d}{2}}\sqrt{\frac{Q^{(3)}}{R^{(3)}}}X(\gamma p,\gamma^2 t;\gamma^{d-2}R,\gamma^{\frac{d-4}{2}}g,\gamma^{-2}Q,\gamma^{-2}M,\gamma^{-2-\frac{d}{2}}b,\gamma^{-1}\lambda).
\nonumber
\end{equation}
Note that although $\gamma$ is dimensionless, the exponents in the arguments reveal the momentum dimension of the corresponding scaled variable. For example, terms $\gamma p$, $\gamma^{\frac{d-4}{2}}g$ and $\gamma^{-1}\lambda$ tell us that $p$, $g$ and $\lambda$ have momentum dimensions $-1$, $\frac{4-d}{2}$ and $1$, respectively.

\subsection{Tadpoles}

The two main approaches to deal with the critical points for Model II with local interactions are (i) identify the critical point via divergence of the correlation function \cite{Janssen2005, Tauber2014, Tauber2005} or (ii) implementation of an additive field transformation centered about the mean density $X_p = \braket{\psi_p}$ \cite{Janssen1999}. Although the end product is the same, the second method is (subjectively speaking) perhaps a bit more transparent and so adopted for the non-local version of Model II. Thus the following field transformation is defined,
\begin{eqnarray}
\bar{\phi}_p & = & \bar{\psi}_p, \nonumber \\
\phi_p & = & \psi_p - X_p.
\nonumber
\end{eqnarray}
Note that as $X_p = \braket{\psi_p}$ is the mean density, we have $\braket{\phi_p} = 0$.

Then writing action $S = S_0+S_1+S_I$, where $S_0$, $S_1$ and $S_I$ respectively represent quadratic, linear and (cubic or quartic) interaction terms, after some algebraic manipulation,
\begin{align}
S_0 = & -\int_t\int_p \bar{\phi}_p\left[\partial_t-D(\nabla^2-M-gX)\right]\phi_p
\nonumber \\
& + D\int_t\int_p\int_q \bar{\phi}_p\left[Q\hat{Q}_{pq}-gX\hat{R}_{pq} \right] \phi_q
+D\int_t\int_p\int_q \bar{\phi}_p\left[ gX\hat{Q}_{pq}-\frac{R}{2}X^2\hat{R}_{pq}\right] \bar{\phi}_q,
\nonumber \\
S_1 = & -D\int_t\int_p \bar{\phi}_p \left[ (M-Q)X+gX^2-b \right],
\nonumber \\
S_I = & Dg\int_t \int_p \int_q \bar{\phi}_p\phi_p
\left[\hat{Q}_{pq}\bar{\phi}_q-\hat{R}_{pq}\phi_q \right]
-DR\int_t \int_p \int_q \bar{\phi}_p\phi_p\bar{\phi}_q
\left[ X\hat{R}_{pq} \right]
-\frac{DR}{2}\int_t \int_p \int_q \bar{\phi}_p\bar{\phi}_q\phi_p\phi_q
\left[ \hat{R}_{pq} \right].
\label{MCA}
\end{align}

Note, firstly, that for local interactions, this reduces to the expression given in \cite{Janssen1999} once the irrelevant parameter $R$ is sent to zero (which will be seen to happen under rescaling). As we are considering steady state, the mean density $X_p=X$ is independent of space (as is the initial condition) and time, and so can be treated and written as a parameter.

\begin{figure}[t!]
\setlength{\unitlength}{0.1\textwidth}
\begin{picture}(8.5,6.5)
\put(0.00,0.00){\includegraphics[width=16cm]{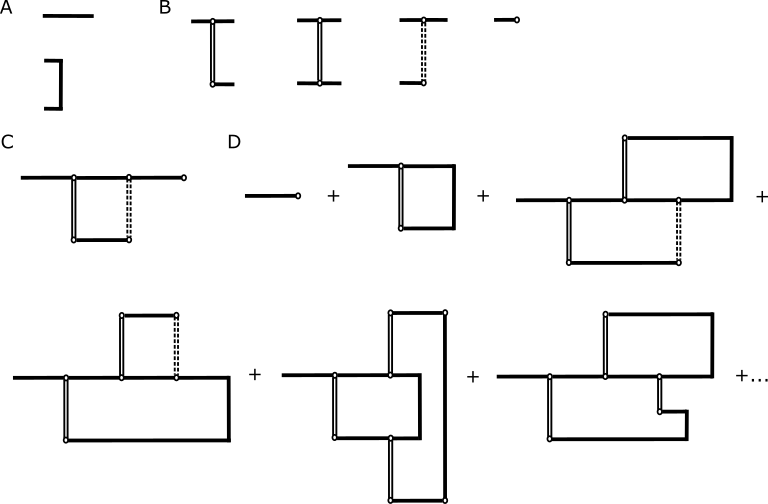}}
\put(0.60,5.92){{\small $G_{\bar{\phi}\phi}$}}
\put(0.80,4.95){{\small $G_{\phi\phi}$}}
\put(6.32,5.71){{\small $b-(M-Q)X-gX^2$}}
\put(5.10,5.31){{\small $D(g\hat{Q}-RX\hat{R})$}}
\put(5.70,5.95){{\small $(k=0)$}}
\put(3.85,5.30){{\small $-\frac{1}{2}DR\hat{R}$}}
\put(2.57,5.30){{\small $-Dg\hat{R}$}}
\end{picture}
\centering
\caption{Feynman diagram elements for the mean centred action in Eq. (\ref{MCA}), including A) the propagators, and B) the internal and initiating nodes. Time runs from right to left. D) The first six elements (up to second loop order) for the tadpole expansion of $\braket{\phi}=0$. C) A tadpole with extra tail.}
\label{TadpolePic}
\end{figure}
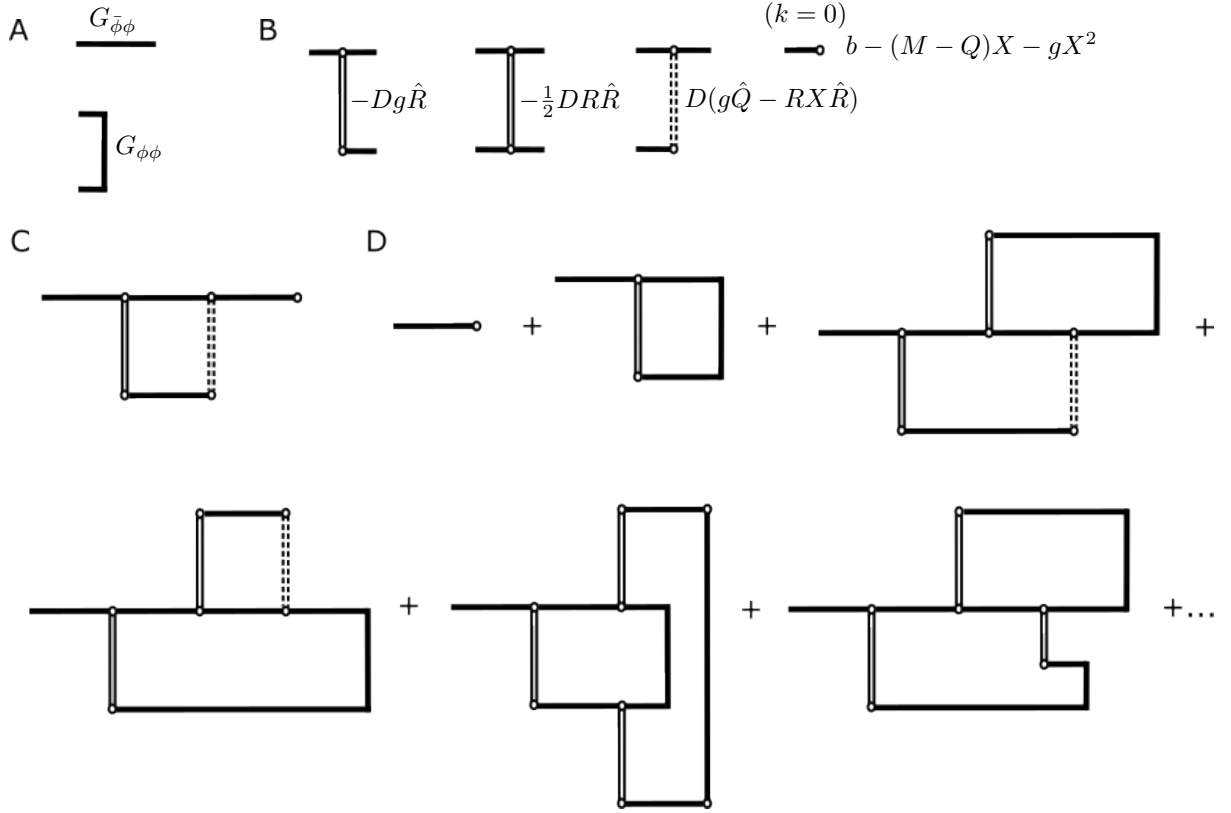

Now, secondly, for Feynman diagrams, the quadratic action terms gives rise to two types of propagator. Deriving these in position space is complicated. In momentum space, it is somewhat more tractable, resulting in the following forms,
\begin{align}
G_{\bar{\phi}\phi}(k,\omega) & = \frac{1}{-i\omega+DF(k,\lambda)},
&
G_{\phi\phi}(k,\omega) & = \frac{2DgX\hat{Q}(k,\lambda)}{\omega^2+D^2F(k,\lambda)^2},
\nonumber
\end{align}
where
\begin{equation}
F(k,\lambda) = k^2+M-Q\hat{Q}(k,\lambda)+gX\hat{R}(k,\lambda)+gX.
\nonumber
\end{equation}
In momentum-time domain these become,
\begin{align}
G_{\bar{\phi}\phi}(k,t) & = \theta(t)e^{-DF(k,\lambda)t},
&
G_{\phi\phi}(k,t) & = \frac{gX\hat{Q}(k,\lambda)}{F(k,\lambda)}e^{-DF(k,\lambda)|t|}.
\nonumber
\end{align}
The linear and interaction terms are then treated perturbatively. The resulting terms can be found in Fig. \ref{TadpolePic}A.

Thirdly, we have the condition that the mean $\braket{\phi_p}=0$. Given that the $n_0$ term has been dropped, there are no initiating nodes at time $t=0$. Together, this tells us that the sum of all tadpole diagrams is zero. The first six diagrams are given in Fig. \ref{TadpolePic}D. Note the tadpoles have no extruding arms such as the diagram given in Fig. \ref{TadpolePic}C, which has an extra edge protruding to the right. If we replace this edge by all the tadpole terms, we have a sum of diagrams including this one. However, they must sum to zero as they have the sum of tadpoles (which is zero) as a factor. Thus any tadpoles that are not 1PI can be dropped (without having to define an effective action) \cite{Peskin1995}. Note that as the particle density is uniform across space, the external momentum in diagrams is zero, meaning the only non-zero momenta present are circulating internal loops. 

Now for local interaction, this results in the following (see \cite{Janssen1999} for details),
\begin{equation}
b = X\left(\bar{\tau}-gX-\frac{4g^2\bar{\tau}^{1-\frac{\epsilon}{2}}}{(4\pi)^{\frac{d}{2}}}
\frac{\Gamma\left(1+\frac{\epsilon}{2}\right)}{\epsilon(2-\epsilon)}
+\frac{2g^4}{3}\bar{\tau}^{-\epsilon}(gX(1-\epsilon)I_1+(2I_1+3I_2)\bar{\tau}))+\dots
\right),
\label{LocalEOS}
\end{equation}
where now $\epsilon=d_c-d=4-d$, and following \cite{Janssen1999}, the parameter definitions below have been adopted ($I_1$ and $I_2$ are numerical constants \cite{Janssen1999}),
\begin{align}
\tau & = M-Q, &
\bar{\tau} & = \tau + 2gX.
\nonumber
\end{align}
Note that the term with the $\epsilon(2-\epsilon)$ denominator (and so divergences at dimensions $2$ and $4$) comes from the single loop diagram in the expansion of Fig. \ref{TadpolePic}D. This has a primitive divergence dimension of two, meaning that additive and field renormalisation are required to raise this above the critical dimension of four.

Note that if we compare the expression in Eq. (\ref{LocalEOS}) to the mean field equation of Eq. (\ref{RiccSS}), we can see a shift in the critical point. Taking the simpler case of  $b=0$ we see that the single loop diagram gives a first order approximation to this shift.

Consider how this shift depends on the form of interaction. Take the first contribution (arising from the single loop diagram). Now for general interaction this takes the following form (the leading arm in all diagrams is divided out in Eq. (\ref{LocalEOS}) and so is dropped),
\begin{equation}
I = -Dg^2X\fint_k \frac{\hat{Q}(k,\lambda)\hat{R}(k,\lambda)}{F(k)}.
\label{SingleLoopDiag}
\end{equation}
For the case of local interaction with $\hat{R}(k,\lambda)=\hat{Q}(k,\lambda)=1$ this reduces to that seen in \cite{Janssen1999}, meaning $F(k,\lambda)$ is quadratic and the primitive degree of divergence is $d=2$ as expected.

Now for non-local interactions, this UV divergence can be muted. For example if we pick normal interactions for one of $\hat{Q}$ and $\hat{R}$, the term in Eq. (\ref{SingleLoopDiag}) will be convergent in any dimension. However, if we pick $\hat{Q}$ and $\hat{R}$ to be local and screened Poisson (in either order), the naive divergence dimension will increase to four and UV divergence is still present at the critical dimension. Thus non-local interactions may not full remove UV divergences.

For the case of local interaction, note from Eq. (\ref{LocalEOS}) that the order $\mathcal{\bar{\tau}^{-\epsilon}}$ term implies divergence near the critical point $\bar{\tau}\rightarrow 0$ (for $\epsilon>0$). This emerges as a result of the last two diagrams in Fig. \ref{TadpolePic}D (see appendix of \cite{Janssen1999} for details). This phenomenon can also be seen to occur for non-local interactions. For example, consider the penultimate diagram in \ref{TadpolePic}D, which evaluates to the following ($\lambda$ dependencies such as $\hat{R}(k) \equiv \hat{R}(k,\lambda)$ are not shown to ease notation),
\begin{align}
I & = -2(Dg)^3\int_{0<t_2<t_1<t}\fint_k\fint_l \hat{R}(k)\hat{R}(l)^2
G_{\bar{\phi}\phi}(k,t-t_2)^2G_{\bar{\phi}\phi}(k,t_2-t_1)
G_{\phi\phi}(k-l,t_2-t_1)G_{\phi\phi}(l,t_2-t_1)
\nonumber \\
& = -2D^3g^5X^2\fint_k\fint_l
\frac{\hat{R}(k)\hat{R}(l)^2\hat{Q}(k)\hat{Q}(k-l)}{F(k)F(k-l)}
\int_{0<t_2<t_1<t}e^{-2DF(k)(t-t_2)}e^{-D[F(k)+F(k-l)+F(l)](t_2-t_1)}
\nonumber \\
& = -2Dg^5X^2\fint_k\fint_l
\frac{\hat{R}(k)\hat{R}(l)^2\hat{Q}(k)\hat{Q}(k-l)}
{F(k)F(k-l)2F(k)[F(k)+F(k-l)+F(l)]}
\nonumber \\
& = -2Dg^5X^2\frac{S_d^2}{(2\pi)^{2d}}\int_0^\infty \int_0^\infty \dd k \hh \dd l \hh
\frac{k^{d-1}l^{d-1}\hat{R}(k)\hat{R}(l)^2\hat{Q}(k)\hat{Q}(k-l)}
{F(k)F(k-l)2F(k)[F(k)+F(k-l)+F(l)]}.
\nonumber
\end{align}
Note that the time integral in the second line is a Laplace convolution over three terms, the two exponentials, and the (unwritten) value $1$, viewed as a (constant) function of first time interval $[0,t_1]$. This gives three poles in momentum-Laplace space. Inversion via the Bromwich integral utilises the pole $\frac{1}{s}$ from the term $1$ to give the denominator in the third line. The other two poles from the exponential terms result in expressions that, for steady state, and so large time, become negligible. The term $S_d$ in the last line represent the surface of a unit radius $d$ dimensional sphere.

Now to see that this is still divergent as we approach the critical point, we first write the factors $\hat{R}(k,\lambda) = 1+\hat{R}_\delta(k,\lambda)$ and $\hat{Q}(k,\lambda) = 1+\hat{Q}_\delta(k,\lambda)$. This means the factor $F(k)$ from the propagators can be written as,
\begin{equation}
F(k) = k^2+M-Q\hat{Q}(k,\lambda)+gX\hat{R}(k,\lambda)+gX = k^2 + \bar{\tau}
-Q\hat{Q}_\delta(k,\lambda)+gX\hat{R}_\delta(k,\lambda).
\nonumber
\end{equation}

Next then we make the transforms $k=\bar{\tau}^{\frac{1}{2}}k'$ and 
$l=\bar{\tau}^{\frac{1}{2}}l'$. Then with the aid of Eq. (\ref{IntSc}), we find,
\begin{equation}
F(\bar{\tau}^{\frac{1}{2}}k') = \bar{\tau}[k^2 + 1
-Q\bar{\tau}^{-1}\hat{Q}_\delta(k,\lambda\bar{\tau}^{-\frac{1}{2}})+gX\bar{\tau}^{-1}\hat{R}_\delta(k,\lambda\bar{\tau}^{-\frac{1}{2}})].
\nonumber
\end{equation}
Now, for the interactions in Table \ref{Interactions}, one can check that that as $\bar{\tau}\rightarrow 0$ the last two terms in this equation are finite. Thus, collecting the resulting powers of $\bar{\tau}$, we find that the penultimate diagram of Fig. \ref{TadpolePic}D is of order $\mathcal{O}(\bar{\tau}^{-\epsilon})$ and we find that even if non-local interaction regulates UV divergence, we still have the same order of divergence near the critical point as $\bar{\tau}\rightarrow 0$ which needs to be dealt with.


\subsection{Renormalisation}

The approach taken then, is as follows. The local interaction renormalisation scheme of \cite{Janssen1999} is adopted. We then use the rescaling methods of Section \ref{AnnRenSec} to approach the critical point, where again we find that solutions to Callan-Symanzik equations can be found directly. Then as we scale towards the critical point, we also find that non-local interactions become local, renormalisation is achieved and the same universality is obtained.

The renormalisation scheme is thus given as follows. We use a subscript $0$ to denote the unrenormalised parameters (that is, the unrenormalised parameters above used thus far). Variables without subscripts are renormalised variables. This is taken from \cite{Janssen1999} and included for convenience; 
\begin{align}
\psi & = Z(u)^{-\frac{1}{2}}\psi_0,&
b & = Z(u)^{-\frac{1}{2}}Z_D(u)b_0,
\nonumber \\
\bar{\psi} & = Z(u)^{-\frac{1}{2}}\bar{\psi}_0, &
D & = Z(u)Z_D(u)^{-1}D_0,
\nonumber \\
X & = Z(u)^{-\frac{1}{2}}X_0, &
\tau & = Z_D(u)Z_{\tau}^{-1}(u)\tau_0,
\nonumber \\
g & = Z(u)^{\frac{1}{2}}Z_D(u)Z_g^{-1}(u)g_0, &
u & = G_{\epsilon}g^2\mu^{-\epsilon}.
\label{RenDef}
\end{align}
Recall that because the mean density $X$ can be viewed as a parameter, independent of space-time in steady state, it can accordingly be renormalised. Note also that the last equation is replacing $g$ with a dimensionless parameter $u$, where $\mu$ is an arbitrary momentum scale, and the term $G_\epsilon = \frac{\Gamma(1+\frac{\epsilon}{2})}{(4\pi)^{\frac{d}{2}}}$, where $\epsilon = d_c-d = 4-d$. The individual factors to order $\mathcal{O}(u^2)$ are given by (again from \cite{Janssen1999}, although note that $g\rightarrow\frac{g}{2}$ and $u\rightarrow\frac{u}{4}$ are needed to get those results, as the actions are defined differently),
\begin{align*}
Z(u) & = 1+\frac{u}{\epsilon}+\left(\frac{7}{\epsilon}-3+\frac{9}{2}\log\frac{4}{3}\right)\frac{u^2}{2\epsilon}, &
Z_D(u) & = 1+\frac{u}{2\epsilon}+\left(\frac{13}{\epsilon}-\frac{31}{4}+\frac{35}{2}\log\frac{4}{3}\right)\frac{u^2}{8\epsilon}, \\
Z_\tau(u) & = 1+\frac{2u}{\epsilon}+\left(\frac{1}{\epsilon}-\frac{5}{16}\right)\frac{8u^2}{\epsilon}, &
Z_g(u) & = 1+\frac{4u}{\epsilon}+\left(\frac{5}{\epsilon}-\frac{7}{4}\right)\frac{4u^2}{\epsilon}.
\end{align*}

Thus, the following implicit equation connects $g_0$ to the renormalized interaction $g$ (and its dimensionless partner $u$),
\begin{equation}
g = g_0Z^{\frac{1}{2}}(u)Z_D(u)Z_g(u)^{-1} = g_0\left(1-\frac{3u}{\epsilon}+\mathcal{O}(u^2)\right).
\label{gRform}
\end{equation}
Under the rescaling of Eq.s (\ref{Rescaling2}) and (\ref{ParamShift2}), and adopting notation such as $g(\gamma)$, for example, to indicate scaled parameters dependence on $\gamma$, with $g=g(1)$ for unscaled parameters, we find,
\begin{equation}
g(\gamma) = \gamma^{-\frac{\epsilon}{2}}g_0Z(u(\gamma))^{\frac{1}{2}}Z_D(u(\gamma))Z_g(u(\gamma))^{-1} = 
\gamma^{-\frac{\epsilon}{2}}g_0\left(1-\frac{3u(\gamma)}{\epsilon}+\mathcal{O}(u(\gamma)^2)\right).
\label{gRgammaform}
\end{equation}

Now, if the squared ratio of Eq.s (\ref{gRform}) and (\ref{gRgammaform}) is taken, the bare parameter $g_0$ can be removed and we find, to second order (using Eq. (\ref{RenDef})),
\begin{equation}
\frac{u(\gamma)}{1-\frac{6u(\gamma)}{\epsilon}+\mathcal{O}(u(\gamma)^2)} = 
\frac{\gamma^{-\epsilon}u}{1-\frac{6u}{\epsilon}+\mathcal{O}(u^2)},
\nonumber
\end{equation}
which rearranges to give,
\begin{equation}
u(\gamma) = \frac{u\epsilon\gamma^{-\epsilon}}{\epsilon - 6u+6u\gamma^{-\epsilon}}.
\label{gRsoln}
\end{equation}
Now we see that (for $\epsilon>0$) as $\gamma\rightarrow 0$, $u(\gamma)\rightarrow \frac{\epsilon}{6}$, and as $\gamma\rightarrow \infty$, $u(\gamma)\rightarrow 0$. These limits flip for $\epsilon<0$. These are the fixed points reported in \cite{Janssen1999}. Either way, with $\gamma=1$ we recover the original parameter $u$.

We can also recover standard Wilson like functions. For example, if we square Eq. (\ref{gRgammaform}) and differentiate,
\begin{eqnarray}
\gamma\frac{\partial u(\gamma)}{\partial \gamma} 
= \frac{\partial u(\gamma)}{\partial\log \gamma} & =
-\epsilon u(\gamma)+u(\gamma)\frac{\partial}{\partial \log\gamma}\left(1-\frac{6u(\gamma)}{\epsilon}\right)
= -\epsilon u(\gamma)-\frac{6u(\gamma)}{\epsilon}
\frac{\partial u(\gamma)}{\partial\log \gamma},
\nonumber
\end{eqnarray}
leading to an equation (to second order),
\begin{equation}
\gamma\frac{\partial u(\gamma)}{\partial \gamma} =
\frac{\partial u(\gamma)}{\partial\log \gamma}
= -\epsilon u(\gamma) + 6u(\gamma)^2.
\nonumber
\end{equation}
The fixed points of $u(\gamma)=0$ or $\frac{\epsilon}{6}$ are again transparent.

These kind of manipulations can be used on the other parameters. For example, for the parameter $\tau$ we have renormalisation equation and scaled versions given by,
\begin{align}
\tau & = \tau_0Z_D(u)Z_\tau(u)^{-1}, &
\tau(\gamma) & = \gamma^{-2}\tau_0Z_D(u(\gamma))Z_\tau(u(\gamma))^{-1},
\nonumber
\end{align}
which leads to the general formula for the rescaling of $\tau$,
\begin{equation}
\tau(\gamma) = \gamma^{-2}\tau\frac{Z_D(u(\gamma))}{Z_D(u)}\frac{Z_\tau(u(\gamma))^{-1}}{Z_\tau(u)^{-1}}.
\nonumber
\end{equation}
In the vicinity of the critical point with small $\gamma$, we find, differentiating,
\begin{equation}
\gamma\frac{\partial \tau(\gamma)}{\partial \gamma}
=\tau(\gamma)\left( -2+\frac{\partial}{\partial \log\gamma}
\log(Z_D(u(\gamma))Z_\tau(u(\gamma))^{-1})\right)
=\tau(\gamma)\left(-2+\frac{\epsilon}{4}\right).
\nonumber
\end{equation}
So for small $\gamma$ we find, 
\begin{equation}
\tau(\gamma) = \tau\gamma^{-2+\frac{\epsilon}{4}}.
\nonumber
\end{equation}

The derivation for $X(\gamma)$ is similar, resulting in full, and small $\gamma$, solutions with forms,
\begin{align}
X(\gamma) & = X\gamma^{-\frac{d}{2}}\frac{Z(u(\gamma))^{-\frac{1}{2}}}{Z(u)^{-\frac{1}{2}}},
& X(\gamma) = X\gamma^{-\frac{d}{2}+\frac{\epsilon}{12}}.
\nonumber
\end{align}

Finally, importantly for Model II, consider how $b$ varies with $\gamma$. The renormalized form is,
\begin{equation}
b = b_0Z(u)^{-\frac{1}{2}}Z_D(u).
\nonumber
\end{equation}
Then under the rescaling of Eq. (\ref{Rescaling2}) this becomes,
\begin{equation}
b(\gamma) = \gamma^{-\frac{d}{2}-2}b_0Z(u(\gamma))^{-\frac{1}{2}}Z_D(u(\gamma)). 
\label{qScale}
\end{equation}
Now, again, we can either take the ratio of these two equations, or differentiate the latter with respect to $\gamma$. Taking the ratio,
\begin{equation}
b(\gamma) = \gamma^{-\frac{d}{2}-2}bZ(u(\gamma))^{-\frac{1}{2}}Z_D(u(\gamma))Z(u)^{\frac{1}{2}}Z_D(u)^{-1}, 
\label{bEq}
\end{equation}
which gives an implicit equation for the variation of $b(\gamma)$. As we approach the fixed point $\gamma\rightarrow 0$, the solution to the Callan-Symanzik equation given in \cite{Janssen1999} can be found. More specifically, from Eq. (\ref{bEq}) we find,
\begin{eqnarray}
\gamma\frac{\partial b(\gamma)}{\partial \gamma}
& = & \left(-\frac{d}{2}-2\right)b(\gamma)+b(\gamma)\gamma\frac{\partial}{\partial \gamma}\log(Z(u(\gamma))^{-\frac{1}{2}}Z_D(u(\gamma)))
\nonumber \\
& = & b(\gamma)\left(-\frac{d}{2}-2+\frac{\partial}{\partial \log\gamma}\log\left(1-\frac{u(\gamma)^2}{8\epsilon}\left(\frac{7}{4}-\frac{17}{2}\log\frac{4}{3}\right)\right) \right)
\nonumber \\
& = & b(\gamma)\left(-\frac{d}{2}-2-\frac{u(\gamma)}{4\epsilon}\frac{\partial u(\gamma)}{\partial \log\gamma}\left(\frac{7}{4}-\frac{17}{2}\log\frac{4}{3}\right)\right)
\nonumber \\
& \rightarrow & b(\gamma)\left(-\frac{d}{2}-2+\frac{\epsilon^2}{144}\left(\frac{7}{4}-\frac{17}{2}\log\frac{4}{3}\right)\right).
\nonumber
\end{eqnarray}
Then for small $\gamma$,
\begin{equation}
b(\tau,X,u,\lambda) = \gamma^{2+\frac{d}{2}-\frac{\epsilon^2}{144}\left(\frac{7}{4}-\frac{17}{2}\log\frac{4}{3}\right)}
b(\tau(\gamma),X(\gamma),u(\gamma),\gamma^{-1}\lambda).
\label{bRescale}
\end{equation} 
Implicit in this equation is the value of $\mu$. This just sets the value of the unscaled renormalised parameters (that is, where $\gamma=1$). Now, although $\gamma$ is dimensionless, $\mu$ has the dimensions of momentum. If we use $\mu$ to make the arguments on the right hand side of Eq. (\ref{bRescale}) dimensionless \cite{Barenblatt1996}, we get an expression analogous to Eq. (19) in \cite{Janssen1999}, where, for small $\gamma$,
\begin{equation}
b(\tau,X,u,\lambda;\mu) = \mu^{2+\frac{d}{2}}
\gamma^{2+\frac{d}{2}-\frac{\epsilon^2}{144}\left(\frac{7}{4}-\frac{17}{2}\log\frac{4}{3}\right)}
b(\mu^{-2}\tau(\gamma),\mu^{-\frac{d}{2}}X(\gamma),u(\gamma),\mu^{-1}\gamma^{-1}\lambda;1).
\label{bRescale2}
\end{equation}
Note, however, that this expression also includes a precision parameter and its scaling. Finally, then, we can consider the behaviour near the critical point. Now from the rescaling in Eq.s (\ref{bRescale}, \ref{bRescale2}) we note that as we approach the critical point with $\gamma\rightarrow 0$ the precision diverges and like Model I we find that the non-local interactions become local. At this point the derivation is identical to that in \cite{Janssen1999}, and the same universal behaviour is observed (see \cite{Janssen1999} for details). Note that we have not derived or used the Callan-Symanzik to do this.


\section{Conclusions}

Here we have shown that for non-local particle interactions in some standard reaction diffusion systems, renormalisation gives the same universal asymptotic behaviour as those observed for local interactions. For short timescales, the non-local interaction has the effect of regulating UV divergence and renormalisation is not always necessary. This depends on the strength of decay of the interaction profile for large momentum. These all constitute a form of regulation analagous to those seen using renormalisation techniques, such as where a momentum cutoff is used, for example, which can be viewed as the strongest form of large momentum decay of an interaction profile. The difference here is that the interaction profile is part of the model, whereas momentum cutoffs are generally taken to infinity in renormalisation group approaches, or Wilson shell integration methods techniques. 

A space-time-field rescaling method was implemented, restricted by the requirement that action structure be preserved. The resulting scaling effectively moves along the characteristics seen in solutions to Callan-Symanzik PDEs. However, the methods used here effectively avoid the equation, although the end result is the same. 

The scaling approach we have adopted can also give some context to the choice of rescaling parameters (such as $\kappa$ and $\mu$), which can seem very abstract when first encountered in reaction diffusion systems. Although in \cite{Lee1994} $\kappa$ is defined via a vertex function, in \cite{Janssen1999} $\mu$ it is an arbitrary scaling. Such scaling parameters are often related to measurements in quantum field theory and therefore tend to have a more concrete meaning and can be somewhat easier to comprehend. Here the rescaling of parameters via $\gamma$ with a preserved action gives context to both models in \cite{Janssen1999, Lee1994}.

For this (and other) work the initial spread of particles is taken as uniform, extending this for other initialisations would be a natural question. Investigating other features such as boundary effects, interaction fronts where interactions are non-local would also be of interest.

Finally we note that the case of two interacting particles has been studied here. For many benchmark studies \cite{Cardy1998, Lee1994, Tauber2014}, the case of $k$ interacting particles is considered. It would be natural to attempt an extension this work to those cases, although an interaction function such as $R(r_1, r_2, \dots, r_k)$ would likely be a somewhat complicated exercise, even for spatially homogeneous systems. 


\section*{Acknowledgements}
The author would like to thank Ben Volymer-Lee and Martin Howard for initial discussions. 


{\footnotesize
\bibliographystyle{abbrv} 
\bibliography{refs_DPNL}
}


\end{document}